\documentclass[twocolumn,superscriptaddress,float,prb]{revtex4-1}
\usepackage{graphicx,amsfonts,amssymb,amsmath,hyperref,hypcap,enumerate}
\usepackage{color}
\usepackage{cleveref}

\newcommand{\kk}{\boldsymbol{k}}

\begin{document}
\title{Electron-phonon and electron-electron interaction effects in twisted bilayer graphene}

\author{Sankar Das Sarma}
\author{Fengcheng Wu}
\affiliation{Condensed Matter Theory Center and Joint Quantum Institute, Department of Physics, University of Maryland, College Park, Maryland 20742, USA}


\begin{abstract}
By comparing with recently available experimental data from several groups, we critically discuss the manifestation of continuum many body interaction effects in twisted bilayer graphene (tBLG) with  small twist angles and low carrier densities, which arise naturally within the Dirac cone approximation for the non-interacting band structure.  We provide two specific examples of such continuum many body theories: one involving electron-phonon interaction and one involving electron-electron interaction. In both cases, the experimental findings are only partially quantitatively consistent with rather clear-cut leading-order theoretical predictions based on well-established continuum many body theories. We provide a critical discussion, based mainly on the currently available tBLG experimental data, on possible future directions for understanding many body renormalization involving electron-phonon and electron-electron interactions in the system. One definitive conclusion based on the comparison between theory and experiment is that the leading order 1-loop perturbative renormalization group theory completely fails to account for the electron-electron interaction effects in the strong-coupling limit of flatband moir\'e tBLG system near the magic twist angle even at low doping where the Dirac cone approximation should apply.  By contrast, approximate nonperturbative theoretical results based on Borel-Pad\'e resummation or $1/N$ expansion seems to work well compared with experiments, indicating rather small interaction corrections to Fermi velocity or carrier effective mass.  For electron-phonon interactions, however, the leading-order continuum theory works well except when van Hove singularities in the density of states come into play.
\end{abstract}

\maketitle

\section{Introduction}
Electronic properties of twisted bilayer graphene (tBLG) at low twist angles are of great current interest because of pioneering experiments \cite{Cao2018Super,Cao2018Magnetic} by Cao et al. and follow up experiments from several groups \cite{Dean2018tuning,MIT2018_rho,Polshyn2019,lu2019superconductors,sharpe2019emergent,serlin2019intrinsic,tomarken2019electronic}.  Among the numerous striking experimental findings, the most significant ones are the discovery of carrier density dependent superconducting and insulating states at low temperatures at various fillings of the tBLG moir\'e flatband, a highly resistive linear-in-temperature($T$) resistivity at higher temperatures above the exotic ground states, and intriguing magnetic properties at different carrier densities.  There is as yet no consensus in the literature on the origin of all the observed exotic phenomena in tBLG except for the general agreement that the physics here is controlled by the extremely flat band nature of the system at low twist angles where the Fermi velocity is greatly suppressed leading to very strong interaction effects.  There are strong hints that the system is strongly correlated and therefore, the noninteracting band theory may not even be a good starting point, but in spite of a large number of theoretical papers \cite{Balents2018,Senthil2018,Koshino2018, Kang2018, Liu2018chiral, Dodaro12018,Isobe2018,You2018, Tang2019, rademaker2018charge, guinea2018electrostatic, gonzalez2018kohn, Lin2018,Lado2018,Vishwanath2018origin, Ahn2018failure,Bernevig2018Topology, hejazi2018multiple,  sherkunov2018novel,kang2018strong,Seo2019,lin2019chiral,Heikkila2018, Lian2018twisted, choi2018electron,Wu2018phonon, wu2019phonon,wu2018topological,wu2019identification,wu2019collective,xie2018nature,bultinck2019anomalous,hazra2018upper,xie2019topology,julku2019superfluid,hu2019geometric} on tBLG, no agreement has been reached on the precise nature of the superconducting and insulating ground states of the system. A serious complication in understanding the tBLG physics at this stage is that experiments do not always agree with each other with respect to the details of the various observed phases and their temperature and carrier density scales, indicating the likely role of unknown nonuniversal physics (e.g. disorder, strain, substrates).

The current work deals with interaction effects in tBLG, but with a rather modest and narrow focus, where we confine ourselves entirely to low carrier doping studying interaction physics close to the Dirac point where the Dirac cone approximation should presumably apply.  Our view is that it may not be particularly useful to insist on one global paradigm underlying all tBLG phenomena, and the possibility that different phenomena may arise from different types of interaction effects should be taken seriously.  After all in normal metals, superconductivity (ferromagnetism) arises from electron-phonon (electron-electron) interactions respectively, and insisting on one interaction mechanism to explain it all (as is often done in high-$T_c$ cuprates) may not be the most reasonable approach.  Given that both electron-electron and electron-phonon interaction effects in monolayer graphene (MLG) are well-understood \cite{CastroNeto2009,Peres2010,DasSarma2011}, we ask the extent to which tBLG properties derive from MLG properties using a continuum many body theory perspective within the Dirac cone approximation.  Such a continuum many body theory approach using the linearized Dirac dispersion (and accounting for spin, pseudospin, and valley) as the starting point has had great success in explaining much of the experimentally observed MLG (as well as regular Bernal-stacked bilayer graphene, BLG, without any twist) phenomenology including electron-electron and electron-phonon interaction effects \cite{CastroNeto2009,Peres2010,DasSarma2011}.  Of course, the strong twist angle induced suppression in the Fermi velocity in tBLG compared with MLG enhances all interaction effects drastically, but it is important to ask whether such a continuum Dirac description of tBLG using a suppressed Fermi velocity is capable of capturing the currently observed experimental phenomenology near the Dirac point at least qualitatively as a zeroth order approximation.  If not, then tBLG must be thought of as an independent strongly correlated lattice system on its own which is not adiabatically connected at all to MLG (i.e. highly successful continuum many body theories cannot be simply transplanted from MLG to tBLG).  A natural question would then arise on how and why (and at what critical twist angle) the continuum many body Dirac approximation breaks down as the twist angle decreases since we know definitively that at large twist angles (i.e. in usual MLGs and BLGs) such a Dirac description is often adequate  \cite{CastroNeto2009,Peres2010,DasSarma2011}.

There is a deeper and more fundamental field theoretic purpose underlying the current work.  Graphene, with its chiral linear particle-hole energy dispersion, is an ideal low-energy effective model for quantum electrodynamics (QED), albeit in two dimensions with the nonrelativistic and unretarded  ( graphene Fermi velocity $\sim c/300$, where $c$ is the velocity of light in vacuum) Coulomb interaction.   So, tBLG, with its suppressed Fermi velocity enables (since the coupling constant is inversely proportional to velocity) an effective analog emulation of QED at very large coupling constant, which is impossible in the usual QED with the vacuum fine structure constant being 1/137 at ordinary energies.  Similarly, for the electron-phonon interaction, the effective dimensionless electron-phonon coupling constant, i.e., the Eliashberg constant \cite{eliashberg1960interactions,eliashberg1961temperature}, also goes as the inverse of the Fermi velocity (i.e. effectively as the effetive tBLG fine structure constant, other parameters being constant) enabling a study of the strong-coupling regime of electron-phonon interaction in tBLG, which is impossible in regular untwisted graphene where the Eliashberg constant is very small, thus restricting the system to the extreme weak-coupling regime. Thus, tBLG at low twist angles close to the magic angle, enables a study of continuum many body phenomena (for both electron-electron and electron-phonon interactions) in the extreme strong-coupling regime simply by virtue of the strongy enhanced effective fine structure constant because of the twist-angle induced Fermi velocity suppression, making the subject particularly interesting.

We consider two specific examples where the experimental data are available from multiple groups to carry out our theoretical work.  We compare the leading order Dirac prescription based continuum many body theories with the experimental data to reach some qualitative conclusions.  The two examples we choose pertain respectively to the roles of electron-phonon interaction and electron-electron interaction in affecting the electronic properties.  Of course, we cannot rule out other effects in each case, but our motivation comes from the fact that in each case, the corresponding MLG experimental results can be well-understood by considering only electron-phonon and electron-electron interactions respectively in each case.

The electron-phonon interaction part applies to the recently predicted \cite{wu2019phonon} and observed \cite{MIT2018_rho,Polshyn2019,lu2019superconductors} linear-in-$T$ resistivity in tBLG with large values of both the absolute resistivity and the temperature coefficient of resistivity.  The basic idea propounded in Ref.~\onlinecite{wu2019phonon} is that the effective electron-phonon coupling determining the finite-temperature electrical resistivity increases inversely proportional to the square of the Fermi velocity in Dirac-like materials, and since the tBLG Fermi velocity is very strongly suppressed with decreasing twist angle, the phonon-induced electrical resistivity (and the associated temperature coefficient) would increase strongly in tBLG with decreasing twist angle.  The velocity suppression effect could be quantitatively large as a factor of 50 decrease in the Fermi velocity would correspond to a factor of 2500 increase in the resistivity and the temperature coefficient.  Indeed, the temperature coefficient of resistivity in low twist angle tBLG at higher temperatures, where the resistivity is linear in $T$, is found to be $\sim$ 100 $\Omega$/K \cite{wu2019phonon,Polshyn2019} whereas the corresponding MLG value is typically 0.1 $\Omega$/K \cite{Hwang2008,Kim2010}. In addition, the experimentally observed temperature coefficient of the tBLG resistivity in the linear-in-$T$ regime is approximately independent of the tBLG doping density consistent with the theoretical prediction.
In the current work, we critically investigate the crossover temperature down to which the linearity in the resistivity arising from electron-phonon interaction effect should persist and examine its relevance to new temperature dependent resistivity data which have become available very recently. \cite{lu2019superconductors}  In particular, we discuss whether such a linear-in-$T$ tBLG resistivity could be construed to be associated with strange metallicity and Planckian behavior as has recently been speculated \cite{MIT2018_rho} in contrast to the phonon scattering mechanism predicted in Ref.~\onlinecite{wu2019phonon} which has considerable experimental support \cite{Polshyn2019}.

For the electron-electron interaction part, we address a key question of considerable importance: What is the role of the quantum electrodynamic (QED) type ultraviolet-divergent many-body renormalization of the graphene Fermi velocity due to the Dirac nature of the low-energy spectrum given that the relevant bare fine structure constant ($ \sim 1/v_F$, where $v_F$ is the Fermi velocity) is large ($>$10 compared with $<$1 in MLG) in low angle tBLG by virtue of the moir\'e flatband induced bare velocity suppression?  Naively, one expects a huge interaction-induced increase (decrease) in the renormalized Fermi velocity (coupling constant) at low energies which should dominate physics near the charge neutrality point. (This is the so-called the ``running of coupling constant'' phenomenon well-known from the renormalization group, RG, flow in QED.)  By carefully examining the available experimental data on the density dependent tBLG effective mass from SdH \cite{Cao2018Super,Cao2018Magnetic,Polshyn2019} and capacitance \cite{tomarken2019electronic} measurements, we comment critically on the evidence for or against the existence of the QED-type coupling constant RG flow in the system.  We mention that such renormalization effects have been widely reported in MLG experiments, where the coupling constant is small, using SdH, capacitance, STM, and ARPES measurements (see Ref.~\onlinecite{Barnes2014} for a discussion of the MLG experiments on the MLG RG flow physics).  Since the bare fine structure constant is much larger in tBLG by virtue of much smaller bare Fermi velocities, one expects huge QED-type interaction effects to manifest in tBLG at low energies (i.e. at low densities close to the Dirac point).  We discuss qualitatively and quantitatively the evidence for or against such enhanced interaction effects using continuum field theories.

\begin{figure}[t!]
	\includegraphics[width=1\columnwidth]{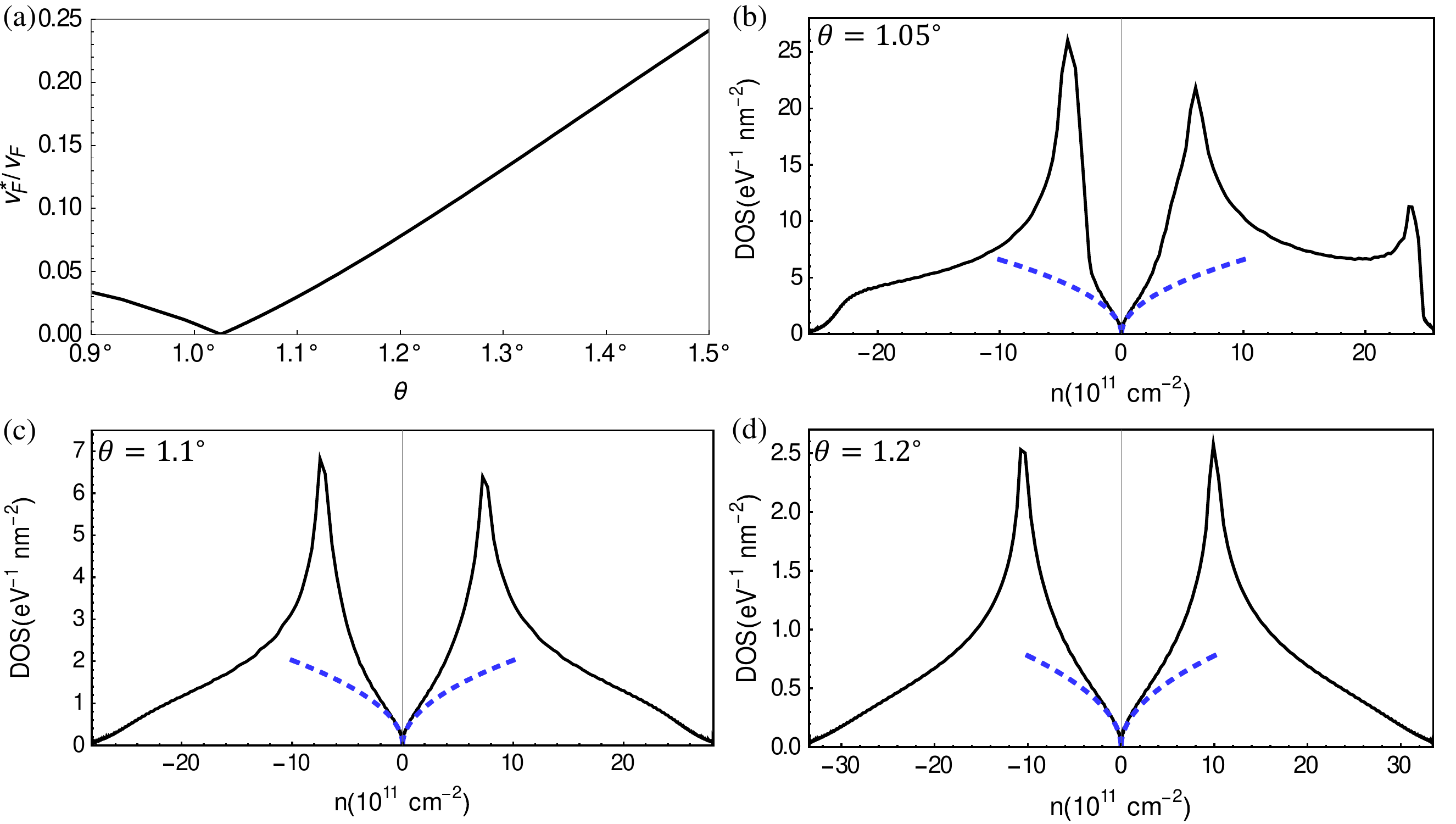}
	\caption{(a) Velocity $v_F^*$ at the Dirac point in tBLG moir\'e bands as a function of twist angle $\theta$. Details of the calculation  can be found in Ref.~\onlinecite{wu2019phonon}. (b), (c) and (d) Density of states (DOS) per spin and valley as a function of total carrier density ($n$). The blue dashed lines show the DOS estimated using the Dirac cone approximation. $\theta$ is respectively 1.05$^{\circ}$ in (b), 1.1$^{\circ}$ in (c)  and 1.2$^{\circ}$ in (d).}
	\label{Fig:vFDOS}
\end{figure}

In order to avoid any confusion or misunderstanding, we emphasize at the outset that the Dirac model of linearly dispersing electron-hole bands used in our continuum approach applies to tBLG only near the charge neutrality point (i.e., low doping densities), and therefore, our theoretical considerations apply to the low carrier density situation with the tBLG chemical potential being below the van Hove singularities of the moir\'e miniband.  We show the density of states for tBLG continuum band structure (at a few values of the twist angle) in Fig.~\ref{Fig:vFDOS} calculated following Ref.~\onlinecite{Bistritzer2011,wu2019phonon} where the linear Dirac cone structure is apparent at low densities.  Our work applies only to this low energy regime which restricts us to a carrier density (electrons or holes) below $\sim 10^{12}$ cm$^{-2}$. The Dirac model does not apply to tBLG above this density, and we have nothing to say about higher doping density situations.  We also show in Fig.~\ref{Fig:vFDOS}(a) the calculated tBLG Fermi velocity at the Dirac point as a function of twist angle.  This Fermi velocity, $v_F^*$, describes the noninteracting low-energy Dirac-like tBLG Hamiltonian in our continuum approximation, which is the starting point of our work. The twist angle dependent tBLG Dirac velocity shown in Fig.~\ref{Fig:vFDOS}(a) is a key parameter in our theory.  We mention that while for MLG the Dirac approximation holds up to $5\times 10^{14}$ cm$^{-2}$ doping density, the same holds only up to a doping of $\sim  10^{12}$ cm$^{-2}$  in tBLG.  

\begin{figure}[t!]
	\includegraphics[width=1\columnwidth]{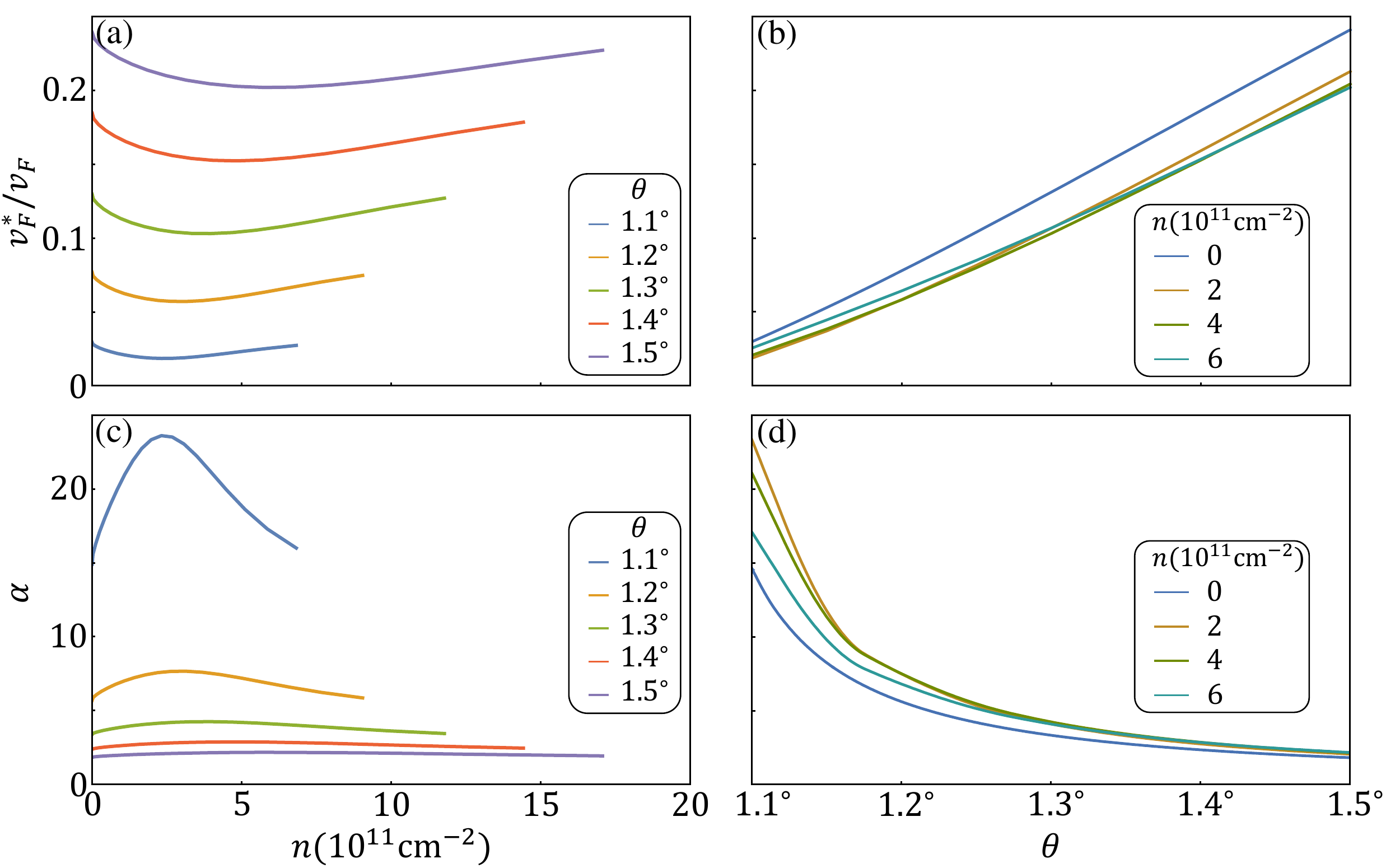}
	\caption{(a) and (b) Velocity $v_F^*$ at finite densities in tBLG moir\'e bands for different twist angles. $v_F^*$ is calculated for the first moir\'e conduction band states along a high-symmetry momentum path that connects corner and center of moir\'e Brillouin zone. In (a), $v_F^*$ is shown for densities below van Hove singularities. (b) and (c) Corresponding values for the effective fine structure constant $\alpha = e^2/(\kappa \hbar v_F^*)$. The dielectric constant $\kappa$ is taken to be 5 for tBLG encapsulated by hBN.}
	\label{Fig:vFalpha}
\end{figure}

For completeness (and because the low density and low twist angle suppression of the tBLG Fermi velocity is the key physics controlling the continuum many body properties of the system near the charge neutrality point), we also show in Fig.~\ref{Fig:vFalpha} the calculated tBLG Fermi velocity as a function of both the doping carrier density and twist angle along with the corresponding effective tBLG fine structure constant (for graphene encapsulated by hBN as used in experimental tBLG systems), which is simply proportional to the inverse of the Fermi velocity. Fig.~\ref{Fig:vFalpha} shows that $v_F^*$ of the non-interacting band structure has a strong angle dependence, but only a weak density dependence at a given angle. We neglect the density dependence of the non-interacting Dirac velocity $v_F^*$ thereafter.  The results shown in Figs.~\ref{Fig:vFDOS} and \ref{Fig:vFalpha}, which present the main physical parameters describing tBLG continuum many body effects at low carrier densities are calculated based on the well-established  band structure model of Ref.~\onlinecite{Bistritzer2011} with parameters given in Ref.~\onlinecite{wu2019phonon}.

The rest of this manuscript is organized as follows.  In Sec.~\ref{sec:eph} we describe the electron-phonon interaction induced tBLG resistivity, discussing how our earlier theory \cite{wu2019phonon} compares with the new transport data \cite{lu2019superconductors} which just became available, specifically commenting on the crossover temperature scale down to which a linear-in-$T$ phonon induced resistivity behavior should hold.  In Sec.~\ref{sec:ee}, we discuss electron-electron interaction effects on the renormalization of tBLG Fermi velocity, and how this renormalization should affect the measurement of the density-dependent tBLG effective mass. In Sec.~\ref{sec:interplay}, we discuss the interplay of electron-phonon and electron-electron interactions.  We conclude in Sec.~\ref{sec:con} with a summary of our main findings as well as a discussion of the open questions and possible future directions.

\section{Electron-phonon interaction}
\label{sec:eph}
We consider electron-acoustic phonon interaction within the deformation potential approximation with the longitudinal acoustic phonons of graphene interacting with the tBLG Dirac electrons in the moir\'e miniband. Assuming the phonons to be unaffected by the tBLG structure (i.e., taking the tBLG phonons to be the same as the MLG phonons), we can write, following Refs.~\onlinecite{wu2019phonon,Hwang2008,Min2011}, the phonon-scattering induced intrinsic carrier resistivity $\rho$ to be given, within the Boltzmann transport theory, by:
\begin{equation}
\rho(T, n, \theta)=\frac{32 F(\theta) D^2 k_F}{ (g_s g_v g_l) e^2 \rho_m v_F^{*2} v_{ph}} I(T/T_{BG})
\label{rhoTntheta}
\end{equation}
where
\begin{equation}
I(z)=\frac{1}{z}\int_0^{1} dx x^4 \sqrt{1-x^2} \frac{e^{x/z}}{(e^{x/z}-1)^2} 
\label{rhoI}
\end{equation}
We use Eq.~(\ref{rhoTntheta}) as it is without derivation, following our earlier work \cite{wu2019phonon} where the details leading to Eq.~(\ref{rhoTntheta}) can be found.

In Eq.~(\ref{rhoTntheta}), $D$, $\rho_m$, and $v_{ph}$ define the phonon model, being respectively the electron-phonon deformation potential coupling constant, the atomic mass density, and the phonon (i.e., sound) velocity. Carriers in tBLG are characterized by $e$, $k_F$, $g_{s,v,l}$, and $v_F^*$, which are respectively the electron charge, the Fermi wave number, the tBLG degeneracy factor (with $g_s$, $g_v$, $g_l$ being each equal to 2 in the absence of any symmetry breaking) arising from spin $(g_s)$, valley $(g_v)$ and layer $(g_l)$ quantum number, and the effective twist angle dependent tBLG Dirac velocity. The Fermi wave number $k_F$ depends on carrier density $n$ through the formula:
\begin{equation}
k_F=\sqrt{4\pi n /(g_s g_v g_l)}.
\end{equation} 

\begin{figure}[t!]
	\includegraphics[width=0.7\columnwidth]{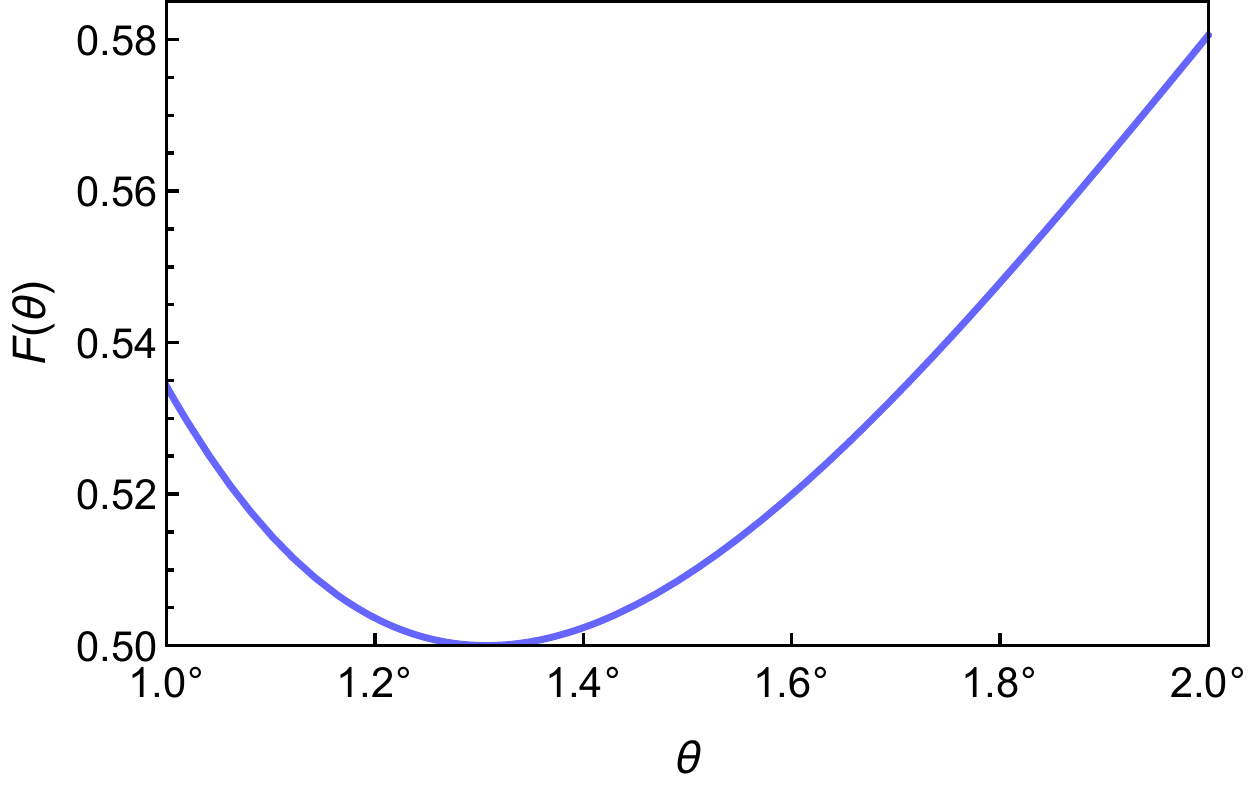}
	\caption{Form factor $F(\theta)$ as a function of the twist angle $\theta$.}
	\label{Fig:F_theta}
\end{figure}

An important physical quantity for our consideration in Eq.~(\ref{rhoTntheta}) is $T_{BG}$, the Bloch-Gr\"uneisen temperature defined by:
\begin{equation}
k_B T_{BG} = 2 \hbar v_{ph} k_F,
\end{equation}
where $k_B$ is the Boltzmann constant. Note that $T_{BG}$ basically defines the energy of an acoustic phonon with a wave number $q=2k_F$. Finally, the function $F(\theta)$ in Eq.(\ref{rhoTntheta}) is a form factor of order unity, which arises from the detailed tBLG moir\'e wave function and accounts for the modification of the tBLG electron-phonon interaction matrix element compared with the MLG situation. Since $F(\theta) \sim 1$ for tBLG according to detailed calculations \cite{wu2019phonon,li2019phonon}, we neglect $F(\theta)$ from qualitative discussions in the following but keep it in the actual quantitative estimation of resistivity. We show in Fig.~\ref{Fig:F_theta} the calculated tBLG electron-phonon form factor $F(\theta)$ as a function of $\theta$ using the tBLG moir\'e band structure of Ref.~\onlinecite{wu2019phonon}. We note the important point that Eqs.~(\ref{rhoTntheta}) and (\ref{rhoI}) apply equally well to regular MLG \cite{Hwang2008,Min2011} except that $F(\theta)=1$, $g_l=1$, and $v_F^*(\theta) \equiv v_F$, where $v_F$ is the regular monolayer graphene Dirac velocity given by $v_F\approx 10^8 $cm/s. In tBLG, $v_F^*<v_F$, because of the moir\'e flatband physics being dominant at low twist angles.

As explained in Refs.~\onlinecite{wu2019phonon,li2019phonon}, the key physics of electron-phonon interaction strength in tBLG compared with MLG or BLG is the strong moir\'e flatband induced enhancement of the effective electron-phonon coupling due to the presence of the $v_F^{*2}$ term in the denominator of Eq.~(\ref{rhoTntheta}). Since $v_F^*(\theta)/v_F \ll 1$ for $\theta \sim \theta_{M}$, where $\theta_{M}$ is the largest magic angle for vanishing Dirac velocity (i.e., $v_F^*(\theta_M)=0$), the phonon-induced tBLG resistivity can be order of magnitude larger than the corresponding MLG resistivity for small twist angle:
\begin{equation}
\rho_{\text{tBLG}}(T, n, \theta) \approx \Big(\frac{v_F}{v_F^*(\theta)}\Big)^2 \rho_{\text{MLG}} \gg \rho_{\text{MLG}}.
\end{equation}
The above physics has already been emphasized in Refs.~\onlinecite{wu2019phonon,li2019phonon} and is consistent with experimental findings \cite{MIT2018_rho, Polshyn2019, lu2019superconductors}. In fact, good quantitative agreement between theory and experiment can be achieved through adjusting the phonon parameter $D/v_{ph}$ by a factor of 2 or so compared with its MLG value, as already discussed in Refs.~\onlinecite{wu2019phonon,Polshyn2019}. Since $D$ is never precisely known, and both $D$ and $v_{ph}$ are likely to be quantitatively modified in tBLG (e.g, by lattice relaxation), the quantitative agreement between theoretical and experimental $\rho(T)$ is quite reasonable. 

The current work focuses on the qualitative temperature dependence, i.e., the power law dependence of $\rho(T, n, \theta)$ on $T$ for different $n$ and $\theta$. We also connect our theory with newly available $\rho(T)$ data in Ref.~\onlinecite{lu2019superconductors}. Expanding the integral in Eq.~(\ref{rhoI}), it is easy to see that
\begin{equation}
\begin{aligned}
&I(z) \sim z^4, \text{for   } z \ll 1, \\
&I(z) \sim z,   \,\,\,\text{for   } z \gg 1,
\end{aligned}
\label{Iz}
\end{equation}
and hence, we get:
\begin{equation}
\begin{aligned}
&\rho(T) \sim T^4, \text{for   } T \ll T_{BG},\\
&\rho(T) \sim T,   \,\,\,\text{for   } T \gg T_{BG}.
\end{aligned}
\label{rhoT}
\end{equation}
Note that the above low-$T$ ($\rho \sim T^4$) and high-$T$ ($\rho \sim T$) [Eqs.~(\ref{Iz}) and (\ref{rhoT})] regimes are known as the Bloch-Gr\"uneisen (BG) and the equipartition regime, respectively, where the phonon scattering induced resistivity is negligible and dominant, respectively. In 3D metals, the $T^4$-BG law  appropriate for 2D systems is replaced by the $T^5$-BG law, and the characteristic temperature scale is the Debye temperature $T_D$. The relevant phonon temperature scale is actually $T_D$ or $T_{BG}$, whichever is lower as shown and discussed in depth in Refs.~\onlinecite{Kawamura1992, Hwang2008,Min2011,Hwang2018}. In low-density electronic materials, the characteristic phonon temperature is $k_B T_{BG}=2 \hbar v_{ph} k_F $ since $T_{BG}<T_D$ when the carrier density is low. In 3D metals, by contrast, $T_D<T_{BG}$, and hence $T_D$ is the relevant temperature scale. Graphene Debye temperature plays no role in the physics of our interest since we are explicitly in the $T_{BG}<T_D$ regime.

Expansion of the integral in Eq.~(\ref{rhoI}) as well as a direct numerical evaluation [see, e.g., Figs. 4(c) and 4(d) in Ref.~\onlinecite{wu2019phonon}] shows that the actual crossover temperature $T_L$, controlling the crossover from the linear-in-$T$ law for $T>T_L$ to the $T^4$-law for $T<T_L$, happens in tBLG for:
\begin{equation}
T_L \approx T_{BG}/8 = \hbar v_{ph} k_F /4 \approx 5 \sqrt{\tilde{n}} K,
\label{TL}
\end{equation}
where $\tilde{n}$ is the carrier density $n$ expressed in units of $10^{12}$cm$^{-2}$. In Ref.~\onlinecite{wu2019phonon}, we estimated $T_L$ by $T_{BG}/4$ in the text, but numerical results presented in Fig. 4 of that paper shows that $T_{BG}/8$ is a better estimation for $T_L$. In this context, it is appropriate to mention that $k_F (\propto \sqrt{n})$ in tBLG is lower than the corresponding $k_F$ in MLG for the same carrier density by a factor of $\sqrt{2}$ (and hence so is $T_{BG}$) because of the additional layer degeneracy $g_l =2$ in tBLG. To obtain the numerical estimate in Eq.~(\ref{TL}) above, we use the MLG value for $v_{ph}=2\times 10^6 $ cm/s.

Before discussing a comparison between the experimental and theoretical $T_L$ values, we first mention two essential restrictions on the applicability of our theory to tBLG: 
(i) the theory is valid only for $\theta>\theta_c$, where $v_F^*(\theta_c)=v_{ph}$, i.e., the theory applies only when $v_F^*>v_{ph}$; (ii) the theory is valid only for $n<n_c$, where $n_c$ is the carrier density up to which the  Dirac cone approximation remains valid in tBLG (actually, the more precise statement is that  the theory is restricted for $E_F(n) < E_c$, defining $n<n_c$, where $E_c$ is the tBLG band energy up to which the linear Dirac cone approximation applies). For $\theta<\theta_c$, intraband scattering by acoustic phonons, which is the only scattering process included in the current theory, vanishes. For $n>n_c$, the chemical potential is above the tBLG Dirac cone regime, where our theory does not apply. Comparing with the tBLG continuum band structure and the standard graphene sound velocity, we find $\theta_c \approx 1.08^{\circ}$, $n_c \approx 10^{12}$ cm$^{-2}$, and we can only compare with experimental data in the $\theta>1.08^{\circ}$ and $n<10^{12}$ cm$^{-2}$ regime. The high density (i.e., $n>10^{12}$ cm$^{-2}$) and the low twist angle (i.e., $\theta<1.08^{\circ}$) regimes are not accessible to the current theory. But the theory should be reasonablly accurate in the $n<10^{12}$ cm$^{-2}$ and $\theta>1.08^{\circ}$ regime, both qualitatively and quantitatively. We note that the exact value of $\theta_c$ depends on details of the Bistritzer-MacDonald model \cite{Bistritzer2011} as well as phonon velocity, and therefore, is not precisely determined. Nevertheless, we expect $\theta_c$ to be close to the magic angle $\theta_M$, which is experimentally found to be around $1^{\circ} - 1.1^{\circ}$. 

\begin{figure}[t!]
	\includegraphics[width=1\columnwidth]{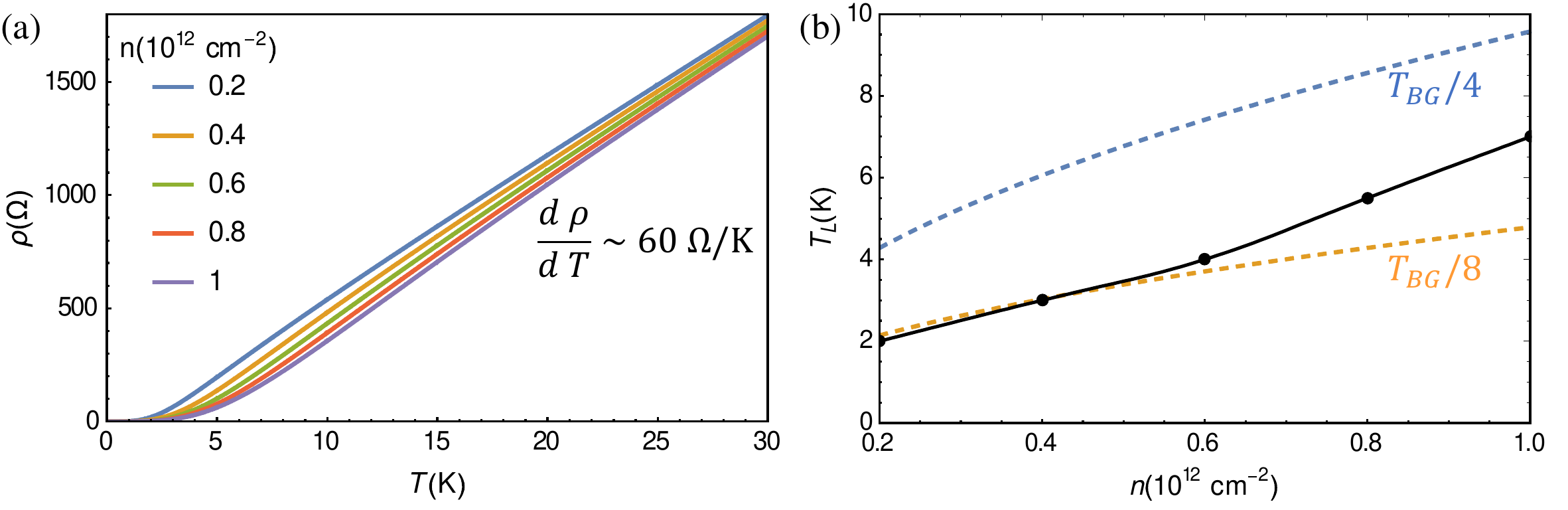}
	\caption{(a) Phonon induced resistivity $\rho$ as a function of temperature $T$ for different carrier density $n$. $\theta$ is 1.1$^\circ$. $\rho$ becomes linear in $T$ at high temperature ($T>T_L$). (b) The crossover temperature $T_L$ as a function of carrier density $n$. Blue and yellow dashed lines respectively show $T_{BG}/4$ and $T_{BG}/8$. }
	\label{Fig:ph_rho}
\end{figure}

First, we summarize the comparison between our theory and the findings in Ref.~\onlinecite{MIT2018_rho,Polshyn2019}, which were already discussed in Ref.~\onlinecite{wu2019phonon}, but some new details are mentioned below:

(1) In both sets of data \cite{MIT2018_rho,Polshyn2019}, where a clear linear-in-$T$ resistivity is seen at lower densities ($<10^{12}$ cm$^{-2}$), where the theory applies, the theory describes the data very well, perhaps even quantitatively if the phonon parameter $D/v_{ph}$ is adjusted upward by a factor of 2. 

(2) In both sets of data \cite{MIT2018_rho,Polshyn2019}, the resistivity is strongly nonlinear at higher densities and lower temperatures, showing that $T_L(n)$ indeed increases with increasing $n$ consistent with the expected $T_{BG} \propto \sqrt{n}$ behavior.

(3) The theory does a poor job of explaining the data quantitatively for  $n>10^{12}$ cm$^{-2}$, most likely because of the failure of the Dirac cone approximation at higher densities.

(4) At lower temperature, typically for $T<10$ K, as well as for the $T=0$ extrapolation from higher temperature resistivity, our theory also does a poor job, partly because of other more dominant (than phonon scatterings) contributions to the low-$T$ resistivity, e.g., impurity scattering \cite{hwang2019impurity} and perhaps also because of fluctuation effects arising from the incipient tBLG superconductivity, which becomes increasingly important at lower temperatures ($<$5 K) for certain carrier densities.

(5) The most important discrepancy between the MIT data \cite{MIT2018_rho} and our theory is for their sample MA4 (Fig. 2a in Ref.~\onlinecite{MIT2018_rho}), where the $T$-linear resistivity for $n=1.19\times 10^{12}$ cm$^{-2}$ and $\theta = 1.16^{\circ}$ persists down to $T\sim 0.5$K just before the system goes superconducting. For $n=1.19\times 10^{12}$ cm$^{-2}$, our theory gives $T_L \approx 5$ K, which is an order of magnitude larger than the experimental $T\sim 0.5$K. In contrast to the very low value of experimental $T_L$ compared with theory, the measured temperature coefficient $d \rho /d T \sim 75 \Omega/$K in this sample is in reasonable agreement with our calculated $d \rho /d T \sim 60  \Omega$/K for $\theta = 1.16^{\circ}$. Of course, $n=1.19\times 10^{12}$ cm$^{-2}$ is above the maximum density $n_c < 10^{12}$ cm$^{-2}$ up to which the Dirac cone approximation (and consequently, our theory) applies, so the persistence of $T$-linearity down to very low temperature may arise from effects beyond our model. In the same figure (Fig. 2a) of  Ref.~\onlinecite{MIT2018_rho}, the authors also show data for their sample MA3 at $n=1.46\times 10^{12}$ cm$^{-2}$ where the linear-in-$T$ behavior persists to $\sim 6$K, which agrees with our theoretical $T_L \sim 6 $K, and therefore, the experimental situation itself is highly nonuniversal, with the experimental $T_L$ values being strongly sample dependent (varying by an order of magnitude), which no universal theory can possibly explain using one resistive mechanism. The persistence of $T$-linear behavior down to very low temperature is even more pronounced in the very recent data of Ref.~\onlinecite{lu2019superconductors}, where only four line plots are provided for $T$-linear resistivity at $n=(0.55, 0.76, 1.11, 1.73)\times 10^{12}$ cm$^{-2}$. The corresponding $T_L$ values according to our theory are respectively, $T_L = 3.7 $K, 4.4 K, 5.3 K and 6.6 K. Experimental $T$-linear resistivity for these four densities persist respectively to temperatures 0.5 K, 0.5 K, 4 K and 6 K. Here, rather unexpectedly, our theory agrees with the regime of $T-$linear behavior for the two higher density ($1.11\times10^{12}$ cm$^{-2}$, $1.73\times10^{12}$ cm$^{-2}$) sample, but not with the two lower density ($0.55\times10^{12}$ cm$^{-2}$, $0.76\times10^{12}$ cm$^{-2}$) samples, although the theory is supposed to apply better to the lower density situation by virtue of the applicability of the Dirac approximation near charge neutrality.  In fact, the discrepancy between theoretical and experimental $T_L$ is again a factor of 10 similar to the MIT MA4 sample at high density \cite{MIT2018_rho}, and similar to the MIT situation, the experimental temperature dependence is nonuniversal with the linearity persisting to 5K in one case and 0.5 K in another for two very similar densities in two different samples. In Fig.~\ref{Fig:ph_rho} we show the calculated $\rho(T)$ and $T_L$, where the discrepancy in $T_L$ for the low-density case compared with the data in Ref.~\onlinecite{lu2019superconductors} is apparent. The persistence of a $T$-linear resistivity down to low temperatures in the low-density samples of Ref.~\onlinecite{lu2019superconductors} is a puzzle for the theory. The discrepancy between experimental and theoretical $T_L$ values in some situations has led the authors of Ref.~\onlinecite{MIT2018_rho} to call the $T$-linear resistivity in tBLG as ``strange metallicity''. We disagree with this characterization as the phonon scattering mechanism does provide an excellent overall description of tBLG $\rho(T)$ for $T>5$K, explaining the overall magnitude of $\rho(T)$ and its roughly density independent large ($\sim 100 \Omega $/K) value for $d \rho /d T$. The simple idea of phonon scattering effect being enhanced by the very large factor $(\sim 10^3 - 10^4)$ of $(v_F/v_F^* (\theta))^2$ in tBLG compared with MLG also provides a natural explanation for why $\rho(T)$ and $d\rho/dT$ are so much (by orders of magnitude) larger in tBLG \cite{MIT2018_rho,Polshyn2019,lu2019superconductors} than in MLG \cite{Kim2010}. The fact that two MIT sample with similar densities (MA3 with $n=1.46\times 10^{12}$ cm$^{-2}$ and MA4 with  $n=1.19\times 10^{12}$ cm$^{-2}$) have very different experimental $T_L$ values (6K and 0.5 K, respectively) is a hint that the $T$-linear behavior in sample MA4 of Ref.~\onlinecite{MIT2018_rho} down to very low $T_L$ may not be a universal phenomenon. A similar experimental discrepancy also applies to Ref.~\onlinecite{lu2019superconductors} where two samples manifest $T_L$ differing by an order of magnitude.

One possibility is that the existence of tBLG van Hove singularities (VHS) for $n>10^{12}$ cm$^{-2}$ is drastically suppressing the effective value of $k_F$ in sample MA4 of Ref.~\onlinecite{MIT2018_rho}, thus reducing $T_L (\propto k_F)$. In fact, the VHS is known to lead to a Lifshitz  transition in the  Fermi surface, providing an effective Fermi wave number corresponding to a much lower carrier density $n_{eff} \approx n-n_{VHS}$, where  $n_{VHS}$ is the carrier density corresponding to the Fermi level being at the VHS. Experiments \cite{MIT2018_rho,Polshyn2019,lu2019superconductors} already show the presence of such small tBLG Fermi pocket for $n> n_{VHS}$. The Fermi wave number $k_{F, eff}$ corresponding to $n_{eff}$ is much smaller than that corresponding to the full density $n$:
\begin{equation}
k_{F, eff} = k_F \sqrt{n_{eff}/n} =k_F \sqrt{(n-n_{VHS})/n}.
\end{equation} 
Given that $T_L \propto T_{BG} \propto \sqrt{n_{eff}}$, $T_L$ could easily be suppressed strongly if $n_{eff} \ll n$. More experimental and theoretical work would be necessary to validate this line of reasoning, but we do believe that a a strong suppression of $T_L$ is possible for $n> n_{VHS}$ because the effective Fermi wave number now correspond to a much smaller carrier density measured with respect to the VHS points because of the Lifshitz transition.

\begin{figure}[t!]
	\includegraphics[width=0.7\columnwidth]{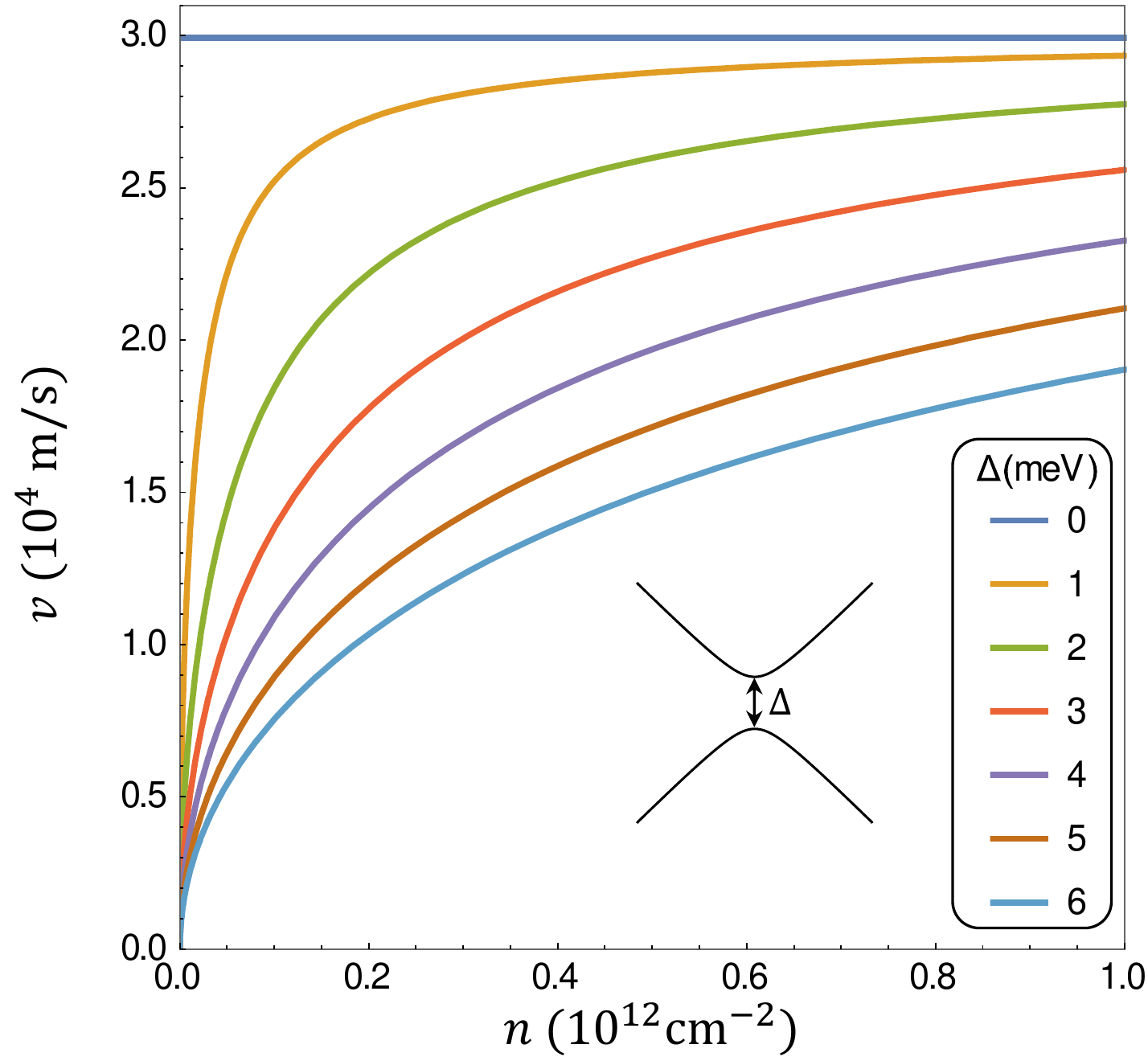}
	\caption{Band velocity $v$ as a function of the total carrier density $n$ (including valley, spin and layer degeneracies) for a massive Dirac Hamiltonian $\hbar v^* \kk \cdot \boldsymbol{\sigma}+ \Delta \sigma_z /2$. Different curves are for different values of the gap $\Delta$, and $v^*$ is taken to be $3\times 10^4 $m/s for this figure.}
	\label{Fig:gapped_Dirac}
\end{figure}

A very recent experiment \cite{he2020tunable} shows a clear correlation between the van Hove singularities in the density of states and the persistence of the $T$-linear resistivity down to rather low temperatures, thus providing considerable support to our proposal of a strong suppression of $T_L$ compared with the nominal $T_{BG}/8$ arising from the Lifshitz transition affecting the effective free carrier density defining $k_F$ in the system.  A full calculation of the tBLG  temperature-dependent resistivity including the Lifshitz transition and the van Hove singularities, which is beyond the scope of the current continuum many body theory, should be carried out in the future to definitively validate the role of the van Hove singularities in extending the $T$-linear esistivity down to $T \ll T_L$.

We note, however, that this van Hove singularity induced Lifshitz transition argument for the suppression of experimental $T_L$ works only at higher density $(>10^{12} $cm$^{-2})$ when the Fermi level is at or above the VHS points. The experimental finding in Ref.~\onlinecite{lu2019superconductors} of very low $T_L$ for $n \approx 0.6 \times 10^{12}$ cm$^{-2}$ is unlikely to be explicable based on an effective lower carrier density arising from the VHS induced Lifshitz transition, although it is not absolutely impossible. A possible explanation for the low-density discrepancy between theoretical and experimental $T_L$ in Ref.~\onlinecite{lu2019superconductors} could be the fact that the sample of Ref.~\onlinecite{lu2019superconductors} are kown to manifest a small energy gap at the charge neutrality point, and hence do not represent a system of massless Dirac fermions at low densities. In fact, close to the charge neutrality point, the system is massive by virtue of the energy gap at the Dirac point. An accurate knowledge of the low-density band structure now becomes crucial for estimating the phonon induced resistivity, and this effect will be the strongest at the lowest density closest to the Dirac point . This might explain why the samples of Ref.~\onlinecite{lu2019superconductors} manifest $T_L$ values which disagree with the theory at lower carrier densities while agreeing at higher densities. 

We show in Fig.~\ref{Fig:gapped_Dirac} the calculated Fermi velocity in a gapped Dirac model  as a function of the energy gap at charge neutrality, and we can see that the effective Fermi velocity is strongly suppressed by the energy gap, particularly at lower carrier density (in fact $v_F^* \rightarrow 0$ as $n \rightarrow 0$). This enhances the effective electron-phonon coupling in these samples considerably (since the coupling goes as $1/v_F^{*2}$) at lower density. Whether the existence of the charge neutrality gap leads to a strong suppressed $T_L$ in sample of Ref.~\onlinecite{lu2019superconductors}  at lower density remains an important open question for the future. Obviously, more work would be necessary to investigate the role of a possible gap at the Dirac point on determining the low-temperature tBLG resistivity.

We emphasize, however, that in the vast majority of the situations studied so far the linearity-in-$T$ in the measured resistivity persists only down to our $T_L$ as defined in Eq.~(\ref{TL}), and hence the continuum many body theory describes the temperature-dependent tBLG resistivity extremely well.  A few cases, where the measured resistivity remains linear well below $T_L$ defined by Eq.~(\ref{TL}), may very well be explicable by the VHS as proposed by us and as recently observed in Ref.~\onlinecite{he2020tunable}

\section{Electron-electron interaction}
\label{sec:ee}

It is well accepted that the massless Dirac fermions in ordinary monolayer graphene manifest the QED-type running of the coupling constant associated with the renormalization group (RG) flow arising from the ultraviolet divergence inherent in the Coulomb coupling of Dirac fermions. The leading-order correction to the effective Fermi velocity arising from electron-electron interactions is easily calculated within the 1-loop approximation to be :
\begin{equation}
v_F^*(E) = v_F^* (E_c) [1+\frac{\alpha_c}{4} \ln (E_c/E)],
\label{vF1loop}
\end{equation}
where $v_F^*(E)$ is the velocity at the energy scale $E$ connecting with the ``bare'' velocity $v_F^*(E_c)$ at high energy scale $E_c>E$. Connecting the whole equation for velocity renormalization to carrier density (using the fact that for Dirac electrons $E_F \propto k_F \propto \sqrt{n}$) we get:
\begin{equation}
\tilde{v}_{F,1}^* / \tilde{v}_{F,2}^* = 1+\frac{\alpha_2}{8} \ln (n_2/n_1),
\label{vFrunning}
\end{equation}
where $\tilde{v}_{F,i}^*$ are the interaction-renormalized Fermi velocity at carrier densities $n_i$ with $n_2>n_1$, and $\alpha_2 = e^2/(\kappa \hbar \tilde{v}_{F,2}^*)$ is the interaction coupling at density $n_2$. Here $\kappa$ is the applicable background dielectric constant of tBLG with the relevant substrates. We assume with no loss of generality that $\tilde{v}_{F,2}^*$ is taken at a sufficiently high carrier density so that it is the twist-angle dependent bare band value:
\begin{equation}
\tilde{v}_{F,2}^* \equiv v_F^*.
\end{equation}
In practical terms, it should suffice to take $n_2=10^{12}$ cm$^{-2}$ where the tBLG band structure at low twist angles starts deviating from linearity making the Dirac cone approximation inaccurate. We note that the characteristic density for the breakdown of the Driac approximation in MLG is very high $\sim 5 \times 10^{14}$ cm$^{-2}$.

The tBLG band Dirac velocity $v_F^*$ (see Fig.~\ref{Fig:vFDOS}(a)) depends on the twist angle $\theta$ and can be empirically approximated by
\begin{equation}
v_F^*(\theta) \approx 0.5 |\theta-\theta_M| v_F,
\label{vF}
\end{equation}
where both $\theta$ and $\theta_M$ are expressed in degrees, $\theta_M \approx 1.02^{\circ}$ from our calculation and $v_F = 10^{8} $ cm/s is the MLG bare Fermi velocity. Equation~(\ref{vF}) holds for $\theta< 3^{\circ}$, and for $\theta>3^{\circ}$, $v_F^* \approx v_F$ for our purpose. Using an effective background dielectric constant $\kappa=5$, which is approximately appropriate for hBN encapsulated graphene systems,  we get $\alpha_{MLG} \approx 0.5$ and $\alpha_{tBLG} = \alpha_2 \approx 0.5 v_F/v_F^*(\theta) \approx 1.0/|\theta-\theta_M|$. Thus, for $\theta = 1.1^{\circ}, 1.2^{\circ}, 1.3^{\circ}, 2^{\circ}$, we have $\alpha_{tBLG}=12, 5.5, 3.6, 1$ respectively. Thus, for small twist angles, the tBLG bare coupling ($\sim 1$) is substantially enhanced compared with MLG  bare coupling by virtue of the strong suppression of tBLG fermi velocity due to the moir\'e flatband situation arising for small twist angles. 

The large bare coupling in tBLG (for $\theta < 2^{\circ}$) should lead to a large increase in the renormalized Fermi velocity $\tilde{v}_F^*$ with decreasing carrier density according to the running of the coupling constant formula in Eq.~(\ref{vFrunning}), giving the following $n$-dependent result:
\begin{equation}
\tilde{v}_{F,1}^*(\theta)/\tilde{v}_{F,2}^*(\theta) = 1 + \frac{\ln (n_2/n_1)}{8|\theta-\theta_M|}.
\label{vFthetarunning}
\end{equation}
Taking $n_2 =10^{12}$ cm$^{-2}$, $n_1 =10^{10}$ cm$^{-2}$, and $\theta= 1.2^{\circ}$, we get $\tilde{v}_{F,1}^*/\tilde{v}_{F,2}^* \approx 4.2$.
The leading order (i.e, 1-loop) RG flow of the coupling constant, therefore, predicts a very large many-body renormalization of the tBLG Fermi velocity as a function of carrier density---going from near the charge neutral point ($\sim 10^{10}$ cm$^{-2}$) to a moderate density ($\sim 10^{12}$ cm$^{-2}$) in a density range where band structure calculations predict, even for a low twist angle of $\theta \sim 1.2^{\circ}$, that the Dirac cone approximation remains valid. We note that the interaction induced renormalization of the Fermi velocity is explicitly density dependent and is always an increasing function with decreasing density, approaching infinite velocity (i.e., zero coupling) at the Dirac point ($n=0$), albeit extremely slowly (i.e., as $\ln n$). Of course, eventually $v_F^*$ must saturate at some exponentially small density at the light velocity $c=3\times 10^{10}$ cm/s ($\gg v_F^*$), where relativistic effects become relevant, but this extreme weak-coupling ($\alpha \approx 1/137$) limit is not of any relevance to the physics of our interest.

\begin{figure}[t!]
	\includegraphics[width=1\columnwidth]{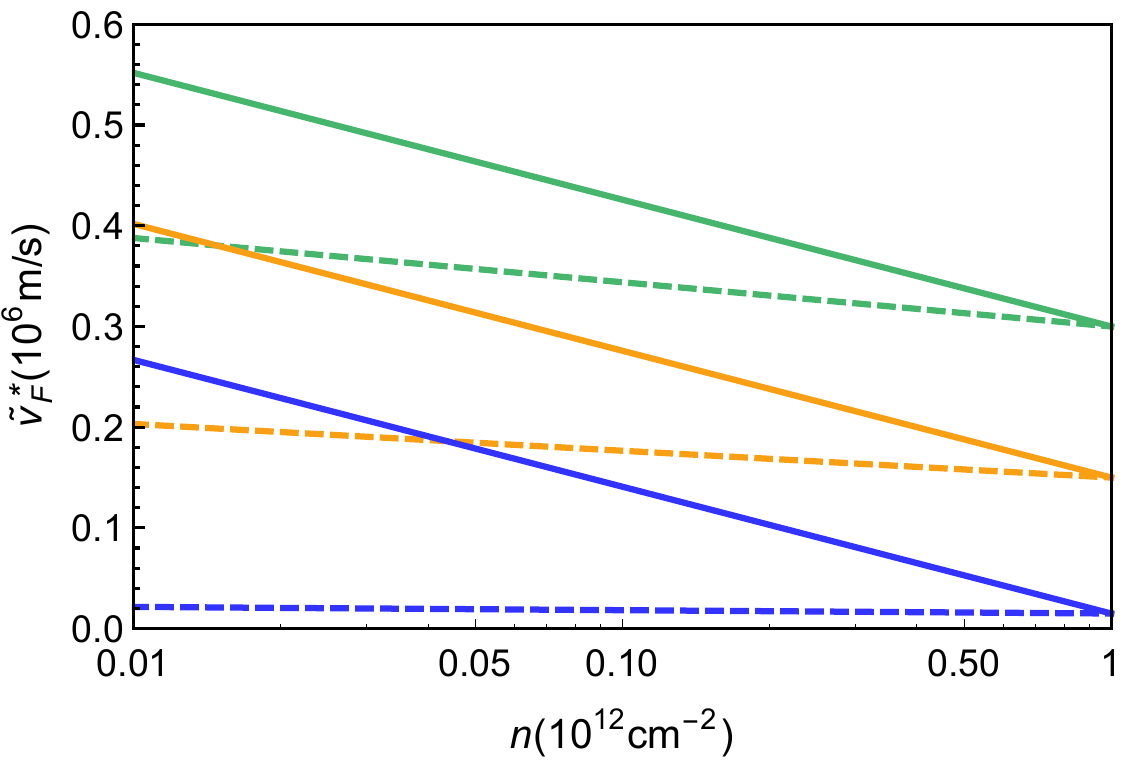}
	\caption{1-loop RG velocity (solid lines) and resummed RG velocity (dashed lines) as a function of total carrier density $n$. The bare velocity is taken to be $0.015 \times 10^6$m/s (blue lines), $0.15 \times 10^6$m/s(yellow lines), and $0.3 \times 10^6$m/s (green lines).}
	\label{Fig:ren_vF}
\end{figure}

To compare with the factor of $\ge 4$ increase in the renormalized tBLG Fermi velocity in going from $n =10^{12}$ cm$^{-2}$ to $n =10^{10}$ cm$^{-2}$ for $\theta=1.2^{\circ}$, we quote the corresponding velocity ratio calculated within the 1-loop theory for the same density variation in regular MLG encapsulated by hBN (where $v_F\approx 10^8$ cm/s and bare $\alpha \approx$0.5): $v_{F, 1, MLG}/v_{F, 2, MLG} \approx 1.3$, where $v_{F, i, MLG}$ corresponds to the effective MLG Fermi velocity for $n =10^{10}$ cm$^{-2} (i=1)$ and $n =10^{12}$ cm$^{-2} (i=2)$. Thus, the coupling constant runs only 30\% in MLG compared with a predicted 420\% running for tBLG with $\theta=1.2^{\circ}$ in the same density range. (As an aside, in high energy physics, OPAL experiments show a maximal QED coupling constant flow only by $\le 3\%$ because of the very weak value of bare coupling, $\alpha \approx 1/137$ in QED \cite{opal2006measurement}.) This 30\% variation in graphene coupling constant has been experimentally observed in MLG by a number of groups using several different techniques (See Ref.~\onlinecite{Barnes2014} and references therein).

It is, therefore, a serious puzzle that the currently available and very recent tBLG experiments find little evidence for a $> 400 \%$ variation in the experimentally measured Fermi velocity as a function of carrier density. In fact, the experimentally extracted Fermi velocity in low twist-angle tBLG typically finds a density independent Fermi velocity which is quite consistent with the noninteracting moir\'e band structure result (for the specific twist angle) in the whole low density regime above charge neutrality ($n<2\times 10^{12}$ cm$^{-2}$), where the Dirac cone approximation should apply. For example, Ref.~\onlinecite{tomarken2019electronic} finds, based on quantum capacitance measurements, $v_F^* \approx 0.116\times 10^8$ cm/s for $\theta =1.05^{\circ}$ (sample M2 of Ref.~\onlinecite{tomarken2019electronic})  over the whole density range $n=0$ to $2\times 10^{12}$ cm$^{-2}$. For $\theta = 1.05^{\circ}$, we have $v_F^*(\theta=1.05^{\circ}) \approx 0.015 \times 10^8$ cm/s from our moir\'e band structure calculation, whereas the corresponding renormalized Fermi velocity including electron-electron interaction effects according to the 1-loop RG theory would be
\begin{equation}
\begin{aligned}
\tilde{v}_F^*(n=10^{11} \text{cm}^{-2}) &\approx 10 \tilde{v}_F^*(n=10^{12} \text{cm}^{-2}) \\
&\approx 0.15\times 10^8 \text{cm/s}.
\end{aligned}
\label{vF1011}
\end{equation}
It is curious that the experiment of Ref.~\onlinecite{tomarken2019electronic} finds a $v_F^* \approx 0.12\times 10^8$ cm/s reasonably comparable to the theoretically expected renormalized Fermi velocity [Eq.(\ref{vF1011})], a factor of 8 larger than the bare band structure value at the density $10^{11} \text{cm}^{-2}$. By contrast, the fact that a constant density-independent Fermi velocity provides description of the data in Ref.~\onlinecite{tomarken2019electronic} argue against the expected running of the coupling constant. One possibility, which cannot be ruled out at this stage, is that the Dirac cone density range covered in the experiment is $10^{10}\sim 2\times 10^{12} \text{cm}^{-2}$, and the data can therefore be reasonably explained by using the renormalized Fermi velocity at the intermediate density of $n\approx 10^{11}\text{cm}^{-2}$, given the error bars in the data. The theoretical RG flow leads to a variation in $\tilde{v}_F^*$ from the bare value of $0.015\times 10^{8}$ cm/s at high density $n\sim 10^{12} \text{cm}^{-2}$ to 20 times the bare velocity (i.e. $0.3\times 10^8$ cm/s) at the lowest density $n \approx 10^{10} \text{cm}^{-2}$, but the noisy experimental data are not precise enough to decisively discern the velocity variation with carrier density. Thus, the data over the whole density range can be fitted approximately by using $\tilde{v}_F^* \approx 0.12 \times 10^8$ cm/s. In Fig.~\ref{Fig:ren_vF}, we show the density dependence of the 1-loop RG velocity. 

\begin{figure}[t!]
	\includegraphics[width=1\columnwidth]{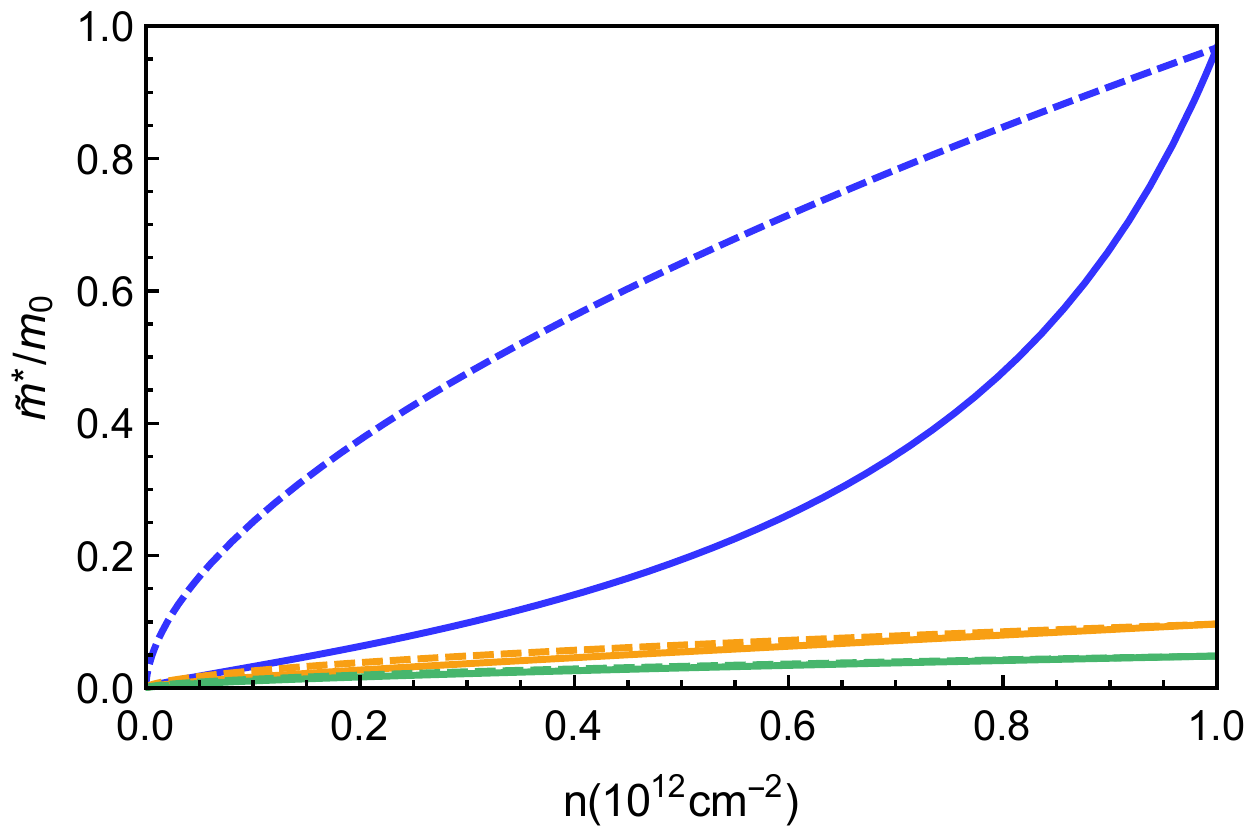}
	\caption{1-loop RG mass (solid lines) and resummed RG mass (dashed lines) as a function of total carrier density $n$. The bare velocity is taken to be $0.015 \times 10^6$m/s (blue lines), $0.15 \times 10^6$m/s(yellow lines), and $0.3 \times 10^6$m/s (green lines).}
	\label{Fig:ren_mass}
\end{figure}

In Ref.~\onlinecite{Cao2018Magnetic}, capacitance measurements on a tBLG sample (device D2 in Extended Data Figure 2 of that paper) with $\theta=1.12^{\circ}$ extracts $\tilde{v}_F^* \approx 0.15 \times 10^{8}$ cm/s in the carrier density range between $10^{10} \text{cm}^{-2}$ and $10^{12} \text{cm}^{-2}$. Here our theoretical bare band velocity $v_F^* \approx 0.05\times 10^8$ cm/s is a factor of 3 smaller than the extracted experimental velocity leading to an effective renormalization factor of 3. Assuming that this renormalization can be modeled by the mid-density $n=10^{11} \text{cm}^{-2}$, we get a factor of 3.9 renormalization arising from the RG flow equation [Eq.~(\ref{vFthetarunning})]. Again, the density independence of the experimental fit to the measured effective velocity could be attributed to the relatively large error bar in the data. 

Thus, the available capacitance based tBLG velocity measurements do not indicate a strong logarithmic running of the coupling constant with decreasing carrier density as predicted by the 1-loop perturbative RG theory.  But the extracted Fermi velocity from capacitance measurements in Refs.~\onlinecite{Cao2018Magnetic,tomarken2019electronic} seems to be much larger than the corresponding theoretical tBLG band velocity, thus suggesting some hints of a velocity renormalization although this is ambigous unless a clear increasing velocity with decreasing carrier density is also observed in the experiment.  We emphasize, however, that the true tBLG bare band velocity may differ from our estimates based on the continuum Bistritzer-MacDonald model, and hence we cannot be sure that the existing capacitance measurements definitively imply any interaction-induced renormalization effects in spite of the 1-loop RG theory predicting a very large velocity renormalization.  The most definitive evidence in favor of the predicted running of the coupling constant would be a direct experimental observation of the logarithmic velocity renormalization, which seems not to have been reported yet.

The theory can also be used to compare with the experimentally measured tBLG effective mass $(\tilde{m}^*)$ through Shubnikov–de Haas (SdH) oscillation experiments in a weak applied magnetic field. Of course, the linear Dirac energy dispersion of graphene implies a vanishing effective mass at the Dirac point corresponding to massless Dirac fermions, but at a finite carrier density away from the Dirac point a graphene effective mass may be defined using the relativistic formula:
\begin{equation}
E_F = \hbar v_F^* k_F  = m^* v_F^{*2},
\end{equation}
leading to:
\begin{equation}
m^*(k_F) =\hbar k_F /v_F^* \propto \sqrt{n}.
\end{equation}
Note that precisely the same effective mass follows also from the Newtonian definition for momentum, i.e., $p=\hbar k_F = m^* v_F^*$, giving $m^*=\hbar k_F/v_F^*$. The so-defined effective mass, which varies at the Dirac point as $\sqrt{n}$, is directly measured from the oscillation amplitude of SdH oscillations. We mention that, for the same value of $n$, the bare tBLG effective mass would be heavier than the MLG effective mass by the large ration of $v_F/(\sqrt{2} v_F^*)$ where the factor of $\sqrt{2}$ arise from the layer degeneracy of tBLG which suppresses its effective $k_F$ by a factor of $\sqrt{2}$ compared with MLG. Using Eq.~(\ref{vF}) for $v_F^*$ in tBLG we get the following formula for the bare tBLG effective mass as a function of twist-angle $\theta$ and carrier density $n$:
\begin{equation}
m^* (n, \theta)/m_0 \approx 0.03 \sqrt{\tilde{n}}/|\theta-\theta_M|,
\end{equation}
where $m_0$ is the electron rest mass and $\tilde{n}=n/(10^{12} \text{cm}^{-2})$. $m^*$ goes to 0 as $\tilde{n} \rightarrow 0$.

This tBLG bare effective mass $m^*$ is renormalized by the coupling constant RG flow due to the velocity renormalization to a many-body effective mass $\tilde{m}^*$ given in the 1-loop theory by:
\begin{equation}
\tilde{m}_1^* (n_1, \theta)= \sqrt{\frac{n_1}{n_2}} \tilde{m}_2^* (n_2, \theta) [1+\frac{\alpha_2}{8} \ln \Big( \frac{n_2}{n_1} \Big)]^{-1},
\end{equation}
where $\tilde{m}_1^*$ and $\tilde{m}_2^*$ are the renormalized effective mass respectively at density $n_1$ and $n_2$ (both for the same system with the twist angle $\theta$). Taking into account the dependence of the bare effective mass already on the density, and taking $n_2 \approx 10^{12} \text{cm}^{-2}$ where the Dirac cone approximation ceases to apply (see Fig.~\ref{Fig:ren_mass}), we get:
\begin{equation}
\begin{aligned}
\tilde{m}_1^*/m_0 \approx 0.03 \sqrt{\frac{n_1}{n_2}} \Big[ |\theta-\theta_M| + \frac{1}{8} \ln \Big(\frac{n_2}{n_1}\Big) \Big]^{-1}
\end{aligned},
\end{equation}
where $n_2$ is taken to be $10^{12} \text{cm}^{-2}$ and $0< n_1 \le n_2 $ so that the Dirac approximation applies.

We can compare the 1-loop effective mass theory to the experimental measurements of the SdH effective mass in Refs.~\onlinecite{Cao2018Super,Cao2018Magnetic,Polshyn2019}. First, we note that the $\sqrt{n}$ dependence of the bare effective mass is suppressed at lower densities in the renormalized effective mass for $n<10^{12} \text{cm}^{-2}$. In fact, as $n \rightarrow 0$, $\tilde{m}^*(n) \sim \sqrt{n}/\ln (n_2/n) \rightarrow 0 $, which varies faster than $\sqrt{n}$ and includes a non-analytic $\ln (1/x)$ function due to the logarithmic RG flow of the Dirac system. Thus, although the renormalized tBLG effective mass $\tilde{m}^*(n)$ at high density $(n \ge 10^{12} \text{cm}^{-2})$ is still given by the bare band mass (since we assume the RG flow to stop at $n \approx 10^{12} \text{cm}^{-2}$ where the Dirac model stops being valid), the low-density effective mass near charge neutrality ($n<10^{12} \text{cm}^{-2}$) should be much smaller than the corresponding bare band mass ($\sim \sqrt{n}$) at that density. Another way of saying this is that a direct effect of many-body renormalization is that the ratio of the effective mass at a low density $n_1$ compared with that at a high density $n_2$($\ge 10^{12} \text{cm}^{-2}$) should be much smaller than that implied by the simple band structure ratio of $\tilde{m}_1^*/\tilde{m}_2^* = \sqrt{n_1/n_2}$ because of the additional logarithmic RG flow of the coupling constant [i.e., $\tilde{m}_1^*/\tilde{m}_2^* = \sqrt{n_1/n_2}/(1+(\alpha_2/8) \ln (n_2/n_1))$]. The upper density cutoff for the RG flow should be set by the experimental van Hove singularity and is thus, dependent on $\theta$, but typically the cutoff density for the RG flow is $\sim 10^{12} \text{cm}^{-2})$ for small values of $\theta$($> \theta_M$) of interest in tBLG experiments. 

In Ref.~\onlinecite{Cao2018Magnetic}, the measured SdH effective mass varies from $\sim 0.1 - 0.2 m_0$ at $n=  10^{11}$ cm$^{-2}$ to $\sim 0.4 m_0$ at  $n=  10^{12}$ cm$^{-2}$ with a rather large error bar. Here $m_0$ is the electron rest mass. Thus, the ratio of the effective mass at these two densities varies by  a factor of 2-4 (with large error bars), which is comparable to that predicted by the noninteracting $\sqrt{n}$ density dependence without the predicted logarithmic coupling constant running effect. Similarly, in Ref.~\onlinecite{Cao2018Super}, the measured effective mass in sample D2 with $\theta=1.1^{\circ}$  manifests the noninteracting $\sqrt{n}$ density dependence (with very large error bars) with an extracted $v_F^*=6\times 10^6$ cm/s which is approximately consistent again with the bare band velocity with this twist angle, thus indicating (within large error bars) an absence of any substantial density-dependent velocity renormalization as predicted by the 1-loop RG theory.

Ref.~\onlinecite{Polshyn2019} obtains the SdH effective mass for a tBLG sample with $\theta \approx 1.59^{\circ}$, finding that the noninteracting result $m^* \propto \sqrt{n}$ is approximately obeyed in the $10^{11} - 10^{12} \text{cm}^{-2}$ range of carrier density (within error bars). This is crudely consistent with theory since for $\theta \approx 1.59^{\circ}$, the bare tBLG Fermi velocity $v_F^*$ is about $0.3 v_F$, leading to only a factor of 3 enhancement of the effective bare coupling constant $\alpha_{tBLG} \approx 3 \alpha_{MLG} \approx 1.5$ so that the logarithmic RG renormalization factor is only $(1+(1.5/8) \ln (n_2/n_1)) \approx 1.4$, and hence, we expect minimal velocity renormalization in the $10^{11} - 10^{12} \text{cm}^{-2}$ density range for the large-angle sample. For the $\theta \approx 1.24^\circ$ sample, the logarithmic RG factor in the $4\times 10^{11} - 10^{12}$cm$^{-2}$ density range is only $\sim 1.5$ because of the rather narrow density range of the measurement. Hence, within error bars, which are typically large for SdH measurements of effective mass, we do expect the bare tBLG Fermi velocity and the $\sqrt{n}$ dependence to describe the data of Ref.~\onlinecite{Polshyn2019}. Lower density experimental measurement (for $n<10^{11} \text{cm}^{-2}$) should allow a more decisive quantitative comparison between theory and experiment. Obviously, the effective mass measurements of Ref.~\onlinecite{Polshyn2019} are actually more consistent with the velocity renormalization being negligibly small in contradiction with the 40\% velocity renormalization predicted by the 1-loop theory.

Because of the large error bars in these effective mass measurements as well as the absence of accurate data at low densities close to the Dirac point ($n<10^{11}$ cm$^{-2}$) where the mass renormalization is the strongest,  much more data would be necessary for a definitive conclusion.  But, it seems pretty clear that the 1-loop RG theory predicts much stronger tBLG mass renormalization than has been reported in the existing (admittedly limited) experimental data.
If the experimental data are available for $v_F^*$ and $m^*$ in the density range of $10^{10} \text{cm}^{-2} - 10^{12} \text{cm}^{-2}$, where the renormalized tBLG Fermi velocity should change, according to the 1-loop RG theory, by a factor of $\sim 800\%$ in a sample with $\theta \approx 1.2^{\circ}$ (so that the bare $v_F^*$ is only 10\% of the MLG Fermi velocity), the situation would be more convincing, particularly since the corresponding renormalization for hBN encapsulated MLG over the same factor of 100 change in carrier density would only be a paltry $\sim 11\%$.

The apparent inconsistency between the existing experiments and the 1-loop RG prediction of a large density-dependent tBLG velocity renormalization ($\gg$100\% compared with the band structure prediction) brings up the important question regarding the applicability of the 1-loop RG flow equation to tBLG because of the large effective bare tBLG coupling constant $\alpha_{tBLG}$ of small twist angles. Since $\alpha_{tBLG}/\alpha_{MLG} = v_{F, MLG}/v_{F, tBLG} \approx 2/(\theta-\theta_M)$ with $\theta$ measured in degrees, $\alpha_{tBLG}\gg \alpha_{MLG}$ for small $\theta \ge \theta_M$. For example, for $\theta=1.1^{\circ} (1.2^{\circ})$, $\alpha_{tBLG} \approx 12 (5.5)$. A natural question now is whether a 1-loop perturbative RG analysis is meaningful in a situation with a bare interaction coupling $\alpha_{tBLG}>1$. Of course, the standard argument is that the RG flow goes toward weak coupling as the renormalized Fermi velocity increases with decreasing density (i.e., energy) starting at the high density (i.e., from above the Dirac cone energy, which is the natural ultraviolet cutoff in the theory). Thus, the weak-coupling RG flow becomes increasingly a more valid approximation as the density decreases. On the other hand, for example, taking $\theta = 1.1^{\circ}$, where the bare coupling $\alpha_{tBLG} \approx 12$, decreasing carrier density by a decade (e.g., going from $10^{12}$ cm$^{-2}$ to $10^{11}$ cm$^{-2}$), the decrease in $\alpha$ is only by a factor of $\sim 4.5$, so the coupling is still pretty large $\sim 3$. Since this is essentially the currently available experimental density scale, the applicability of the 1-loop perturbative RG becomes dubious. (The situation improves somewhat considerably if the experimental tBLG measurements of Fermi velocity and/or effective mass could be pushed down in density to $10^{10}$cm$^{-2}$ since even for $\theta \approx 1.1^{\circ}$, the  renormalized effective coupling in the 1-loop theory decreases to 1.5 which is still larger than unity.) So the whole  comparison between tBLG theory and experiment based on the 1-loop RG theory should be taken with a large grain of salt. A 1-loop perturbative RG theory should not apply to a situation where the bare coupling is larger than unity-- after all, the great success of QED is based entirely on the accidental fact of the vacuum fine structure constant being very small ($\sim$ 1/137).  So, the disagreement between the 1-loop RG theory and tBLG measurements is simply emphasizing the inapplicability of the 1-loop theory to tBLG with its very low bare Fermi velocity.

Actually, the issue of the large bare coupling (i.e., $\alpha>1$) arises already for ordinary free-standing MLG in vacuum without any substrate, where $\alpha_{vac}=e^2/(\hbar v_F) \approx 2.2$. Even the very extensively studied graphene on SiO$_2$ substrate has a relatively large bare coupling of $\alpha_{\text{SiO}_2} \approx 0.8$ compared with the QED fine structure constant $\alpha\approx 1/137$. Therefore, the applicability of the 1-loop RG theory for MLG is also questionable, and has been questioned in the context of graphene experiments  \cite{Barnes2014, Hofmann2014}. A rather extensive analysis \cite{Barnes2014} shows that the field theory underlying graphene is renormalizable, i.e., only a finite number of logarithmic divergences arise in the theory, but the perturbative expansion in $\alpha$ is asymptotic only perhaps to the first term (or even less). In fact, a literal QED type perturbative expansion up to $O(\alpha^2)$ implies a strong coupling fixed point at $\alpha_c \approx 0.7$, where the Fermi velocity in fact flows to zero, implying infinite coupling. Experiments in MLG, by contrast, agree reasonably well with the 1-loop RG theory, and therefore, the 2-loop perturbative strong coupling fixed point (already for $\alpha<1$) is most likely an artifact of the perturbation theory which fails even more strongly at the second order than the first order. This is consistent with the analysis showing that for $\alpha \sim \alpha_c (\approx 0.7)$, the perturbative expansion is asymptotic only up to $O(\alpha)$ and starts diverging beyond that. Obviously, this problem is far worse for tBLG with small twist angles where $\alpha_{tBLG} \sim 10$, and therefore the perturbation expansion is not convergent at all.

An alternative approach is to use the 2-loop results of Ref.~\onlinecite{Barnes2014} and do an approximate Borel-Pad\'e resummation where the explicit 2-loop results show up as the first two terms (i.e., up to $O(\alpha^2)$) of the resummed RG flow equation. The idea is that the final result to all orders in $\alpha$ including non-perturbative effects (e.g., instantons) is finite, and therefore, a Borel-Pad\'e resummation of the perturbative expansion is a more accurate description of the underlying theory than the term by term asymptotic perturbative series itself \cite{Lipatov1977Lett,Lipatov1977}. 

Such a  Borel-Pad\'e resummed renormalization of the graphene perturbative expansion gives the following nonperturbative RG flow equation:
\begin{equation}
\tilde{v}_F^*(E) = \tilde{v}_F^* (E_c) \Big[ 
1+
\frac{\alpha}{4} \Big\{ 1+\Big(\frac{8}{3} - 2 \ln 2 \Big)\alpha \Big\}^{-1} \ln \frac{E_c}{E}
\Big],
\label{vFnonpert}
\end{equation}
where $\tilde{v}_F^*(E)$ is the renormalized Fermi velocity at energy $E<E_c$ and $\tilde{v}_F^*(E_c)$ is the bare velocity at the ultraviolet cutoff energy scale $E_c$ with $\alpha$ being the bare coupling, i.e., $\alpha \equiv \alpha(E_c)$. Note that this resummed RG flow keeps $v_F^*(E)$ finite for all $\alpha$, and it reproduces the known 1-loop and 2-loop results up to $O(\alpha^2)$. In fact, it agrees with the 3-loop results up to $O(\alpha^3)$ pretty well also. One can think of Eq.~(\ref{vFnonpert}) to be the appropriate RG flow equation in the strong-coupling situation which does not suffer from the artifacts of the loop expansion result in powers of coupling.

Connecting energy to carrier density through the usual substitution of:
\begin{equation}
E \rightarrow E_F \propto \sqrt{n}
\end{equation}
we get (see Fig.~\ref{Fig:ren_vF}):
\begin{equation}
\tilde{v}_F^*(n) = \tilde{v}_F^*(n_c) \Big[ 1+ \frac{(\alpha/8)\ln (n_c/n)}{1+\Big(\frac{8}{3}-2 \ln 2\Big)\alpha} \Big],
\label{vFn_resummed}
\end{equation}
where $\tilde{v}_F^*(n)$ and $\tilde{v}_F^*(n_c)$ are respectively the renormalized Fermi velocity at density $n(<n_c)$ and the bare band Fermi velocity at the highest density $n_c$ up to which the linear Dirac cone approximation holds with $\alpha=e^2/(\kappa \hbar v_F^*)$ and $v_F^* = 0.5 v_F (\theta-\theta_M)$ being the twist-angle-dependent tBLG band Fermi velocity. Once $\tilde{v}_F^*(n)$ is known, the renormalized effective mass $\tilde{m}^*$ is defined in the usual way for graphene: $\tilde{m}^*=\hbar k_F/\tilde{v}_F^*$. Comparing the resummed RG flow theory to the same sets of experimental results in Ref.~\onlinecite{Cao2018Super,Cao2018Magnetic,tomarken2019electronic,Polshyn2019}, we conclude as follows. 

(i) For the $\theta \approx 1.05^{\circ}$ sample in Ref.~\onlinecite{tomarken2019electronic}, the resummed thory gives at $n=10^{11}$cm$^{-2}$ a renormalized Fermi velocity $\tilde{v}_F^* \approx 1.25 v_F^* $. Thus, the tBLG Fermi velocity according to the resummed formula is renormalized only by about $25\%$ in going from $n=10^{12}$ cm$^{-2}$ to $n=10^{11}$ cm$^{-2}$, perhaps explaining why Ref.~\onlinecite{tomarken2019electronic} obtains reasonable agreement between theory and experiment using the bare tBLG Fermi velocity over the whole $10^{11} - 10^{12}$ cm$^{-2}$ density range, since a 25\% variation in the Fermi velocity is well within the error bar of the experimental measurements. We note that, by contrast, the corresponding 1-loop formula predicts a factor of $\sim 10$ increase in the renormalized Fermi velocity in decreasing the density from $10^{12}$ cm$^{-2}$ to $10^{11}$ cm$^{-2}$. Since the experiment of Ref.~\onlinecite{tomarken2019electronic} can be well fitted by a single Fermi velocity in the whole density range, it is clear that the resummed nonperturbative theory is in much better qualitative agreement with the experimental data than the 1-loop theory which predicts an order of magnitude variation in the tBLG velocity over the  $10^{11} - 10^{12}$ cm$^{-2}$ density range.

(ii) In Ref.~\onlinecite{Cao2018Magnetic}, the experimentally extracted Fermi velocity $\tilde{v}_F^* \approx 0.15 \times 10^8$cm/s appear to fit the capacitance data (within large error bars) over a $10^{11} - 10^{12}$ cm$^{-2}$ density range for a sample with $\theta = 1.12^{\circ}$. The resummed theory predicts $\tilde{v}_F^*(n=10^{11}\text{cm}^{-2}) \approx 1.21 v_F^*$, where $v_F^* \approx 0.05 \times 10^8$ cm/s is our calculated bare tBLG band velocity for $\theta = 1.12^{\circ}$. Thus, the variation in the renormalized velocity is only 20\% over the $10^{11} - 10^{12}$ cm$^{-2}$ density range according to the resummed theory. By contrast, the 1-loop theory predicts a large change of the velocity by a factor of 4.  Again, the experimental data are qualitatively much more consistent with the resummed theory than the 1-loop theory.

(iii) Finally, we consider the SdH measurements in Refs.~\onlinecite{Cao2018Super,Cao2018Magnetic,Polshyn2019}, where the experimentally extracted effective mass seems to agree with the simple noninteracting prediction of a $\sqrt{n}$ dependence without any obvious signatures (within error bars) of a strong logarithmic deviation from the $\sqrt{n}$ density dependence of $\tilde{m}^*(n)$. This is qualitatively consistent with the resummed nonperturbative formula in the sense that the resummed theory predicts a much weaker logarithmic renomalization $(\le 25\%)$ than the 1-loop theory ($>100\%$) for small twist angles $\theta<1.5^{\circ}$. On the other hand, the experimentally extracted Fermi velocities themselves appear to be much larger than our calculated bare band velocities obtained from the Bistritzer-MacDonald model. Whether this is due to the band theory being inaccurate or the actual twist angle being larger than the quoted twist angle is unclear at this stage.

Clearly, much more density-dependent experimental measurements for $n<10^{12}$cm$^{-2}$ of effective velocity and effective mass for tBLG sample with low twist angles ($\theta<1.3^{\circ}$) are necessary before any decisive conclusion regarding the relevance of the coupling constant RG flow associated with the Dirac cone renormalization can be reached. At this point, with the availability of rather limited data in the low density ($<10^{12}$ cm$^{-2}$) regime, we have only a tentative conclusion as discussed in depth in this section: The relative density independence of the experimentally extracted Fermi velocity and the approximate $\sqrt{n}$ dependence of the extracted effective mass indicate that tBLG renormalization is more consistent with the nonperturbative resummation theory which predicts rather modest $(<25\%)$ tBLG coupling constant renormalization in the $10^{11} - 10^{12}$ cm$^{-2}$ carrier density range even for $\theta$ as small as $1.1^{\circ}$ where the bare coupling is large ($\sim 10$). By comparison,the 1-loop RG theory (valid technically for $\alpha \ll 1$) predicts a $> 100\%$ variation in the renormalized coupling, which is not experimentally observed.

We note that our finding regarding the applicability of the resummed RG theory (i.e. Eq.~(\ref{vFn_resummed})) rather than the 1-loop RG theory (i.e. Eq.~(\ref{vFrunning})) to tBLG experiments is sufficiently important that we hope that experimental tBLG measurements of Fermi velocity and effective mass will be extended to much lower carrier density ($10^{10}$ cm$^{-2}$ or lower) with better quality data (i.e. lower error bars) so as to validate (or invalidate)  this tentative conclusion.  In particular, tBLG experiments at low carrier density should search for the logarithmic density dependence so that the basic idea of the running of the coupling constant can be validated for a strong-coupling system.  The current tBLG experimental measurements do not see any obvious signature of a logarithmic variation, but this could simply be because of large error bars in the measurements.  If it is definitively established that the measured low density tBLG velocity is basically the bare band velocity with no density dependence, then one must conclude that the continuum Dirac description does not apply to tBLG at any energy (or carrier density), and the system must be described by a moir\'e lattice theory at all scales (in contrast to MLG or BLG where the continuum description is very successful over a wide range of doping density).   In this context it might be useful to tune the electron-electron interaction in tBLG by putting the metal gates rather close to each other where screening by the gates would strongly modify the interaction from the $1/r$ Coulomb interaction to a much weaker interaction (e.g. $1/r^3$) leading to substantial modifications in the many body renormalization corrections.  Thus, in addition to the doping density, the distance of the gates from the tBLG layers could be used as additional tools to tune interaction effects.

The fact that the 1-loop perturbative RG theory of Eq.~(\ref{vF1loop}) fails quantitatively for tBLG even near the Dirac point (and that the corresponding resummed nonperturbative Borel-Pad\'e theory of Eq.~(\ref{vFnonpert}) describes the existing tBLG (and MLG) experimental results well) has profound implications, not only for graphene physics, but also for QED.  In particular, the 1-loop theory of Eq.~(\ref{vF1loop}) has the well-known Landau pole\cite{Landau1954} with the effective coupling diverging at a finite energy scale$\sim E_c \exp (4/\alpha)$, which, for $\alpha \gg 1$, as it is in tBLG near the magic angle, would happen essentially at $E_c$, the ultraviolet cut off scale of the tBLG moir\'e lattice.  By contrast, in QED, the Landau pole happens at the unphysical energy scale way above the Planck scale because of the very small value of the QED fine structure constant ($\sim 1/137$).  Nevertheless, the question remains whether the Landau pole of QED (and of all quantum field theories which are not asymptotically free with the running coupling constant increasing with increasing energy) is an artifact of the perturbative RG theories or implies that all such field theories are basically trivial.\cite{jian2020landau}  The fact, as discussed in the current work in depth, that the existing tBLG experiments produce measured effective Fermi velocities and effective masses in clear agreement (disagreement) with the nonperturbative resummed theory (1-loop perturbative theory) strongly indicate that the Landau pole is an artifact of the 1-loop perturbative RG theory and is not a real physical effect.  Given the large effective tBLG fine structure constant because of the strong band velocity suppression, one would have expected very strong many-body renormalization of the Fermi velocity (and the effective mass) if the 1-loop RG theory is valid.  We show in this work that this does not happen, and the measured velocity/mass near the Dirac point are in fact consistent with the rather small many body renormalization effects predicted by the nonperturbative theory defined by Eq.~(\ref{vFnonpert}). Since the nonperturbative theory implies the nonexistence of a Landau pole [i.e. the coupling constant does not run to infinity at a finite energy scale in Eq.~(\ref{vFnonpert}) as it does in Eq.~(\ref{vF1loop})], we conclude that there is no Landau pole in tBLG, and by extension, there is no Landau pole in QED.

In this context, we may consider an alternative theoretical approach to calculating the tBLG many body effects using the so-called $1/N$ expansion (with $N=4$ in tBLG corresponding to the two valleys and two layers).  Here the expansion parameter is $1/N (<1)$ rather than $\alpha (\gg 1)$, and hence the $1/N$ theory should be quantitatively more reliable in tBLG than the loop expansion perturbation theory. In fact, it turns out that the actual expansion parameter in the graphene theory is $1/(\pi N )$ rather than just $1/N$, making the $1/N$ expansion much more reliable than the loop expansion in the effective coupling constant $\alpha$ in the strong-coupling tBLG limit near magic twist angles. (By contrast, in QED, $\alpha \sim 1/137$, so the loop expansion remains asymptotically accurate up to very high orders in the perturbation theory, as is well-known.)  Following Refs.~\onlinecite{Hofmann2014} and \onlinecite{Son2007Quantum} carried out for MLG, we find that the correction to the Fermi velocity implied by the $1/N$ expansion theory in tBLG (with $N=4$) for $\alpha \sim 10$ is only about $10-30\%$ in contrast to the 1-loop RG theory [i.e. Eq.~(\ref{vF1loop})] which predicts, as discussed above, a Fermi velocity renormalization by $400-500\%$ for $\alpha \sim 10$ !  This $10-30\%$ renormalization is consistent not only with the tBLG experiments, it is also consistent with the nonperturbative Borel-Pad\'e resummed theory which predicts a very similar quantitative renormalization.  We therefore conclude that theory and experiment are in agreement here as long as one uses a strong coupling theory (either Borel-Pad\'e resummation or $1/N$ expansion), and not the 1-loop perturbative RG theory which clearly fails for tBLG at low twist angle because of the strong enhancement of the effective coupling constant.

In fact, the strong-coupling limit, for $\alpha\gg 1$, of the $1/N$ expansion can be obtained analytically ($N=4$ for tBLG with layer and valley degeneracy), giving:
\begin{equation} \label{vFlargeN}
v_F^*(E)= v_F(E_c) (p_c/p)^{\delta},
\end{equation}
where $p$ and $p_c$ are respectively the momentum scales corresponding to energies $E$ and $E_c$ (as, e.g., in Eq.~(\ref{vF1loop})), and $\delta$ is the exponent for the Fermi velocity in the $1/N$ expansion, related to the corresponding anomalous dimension $z<1$, defined by:
\begin{equation} \label{exponent}
\delta=1-z.
\end{equation}
Equations (\ref{vFlargeN}) and (\ref{exponent}) define how the Fermi velocity changes with momentum in the infinite coupling ($\alpha \gg 1$) and large-$N$ limit.  They should be contrasted with Eq.~(\ref{vF1loop}) defining the weak-coupling 1-loop perturbative RG result valid for $\alpha\ll 1$.  The approximate anomalous dimension exponent (or the dynamical exponent) up to $O(1/N^2)$ is given by approximately \cite{Hofmann2014}:
\begin{equation} \label{dimensionz}
z= 1 - 4/(\pi^2  N) - 1/(\pi N)^2. 
\end{equation}
We can convert the momentum scales to carrier density $n$ by the usual substitution of $p=k_F=\sqrt{\pi n/2}$, leading to the conclusion that for a change in density from $n=10^{10}$ cm$^{-2}$ to $n=10^{12}$ cm$^{-2}$, the strong-coupling theory of Eq.~(\ref{vFlargeN}) predicts a Fermi velocity change of only $\sim 40\%$ in the infinite coupling limit whereas the weak-coupling theory of Eq.~(\ref{vF1loop}) predicts a $400\%$ change in the Fermi velocity for $\alpha=10$ !  Experimental results, as discussed extensively above, are in total disagreement with the weak-coupling 1-loop RG theory, but are in agreement with the strong-coupling $1/N$ expansion theory.  We, therefore, believe that electron-electron interaction induced tBLG many body effects at low carrier densities are well-described by the appropriate strong coupling continuum field theory, but not by the 1-loop perturbative RG theory.  Another direct conclusion arising from the agreement of the experimental tBLG data with Eq.~(\ref{vFlargeN}) [or Eq.~(\ref{vFnonpert})] is that Landau poles do not exist in tBLG, and experiments are clearly showing the absence of the Landau pole although the weak-coupling 1-loop theory of Eq.~(\ref{vF1loop}) predicts a Landau pole.

\section{Interplay of electron-phonon and electron-electron interactions}
\label{sec:interplay}

Given that both electron-phonon and electron-electron interactions are strongly enhanced in tBLG at small twist angles by virtue of the strong suppression of the tBLG Fermi velocity, an interesting (and potentially important) question arises about their interplay:  Do they just act independently of each other or is there an interplay leading to new physics?  A detailed answer to this question is rather difficult and way outside the scope of this paper, but it is possible to make some comments,  restricting the discussion of this interplay to the electronic properties being studied in this work, namely, the $T$-dependent resistivity (Sec.~\ref{sec:eph}) arising from electron-phonon interaction and the logarithmic Fermi velocity (and effective mass) renormalization (Sec.~\ref{sec:ee}) arising from electron-electron interaction.  There are other effects where electron-electron and electron-phonon interactions produce interesting mode coupling effects, e.g. plasmon-phonon coupling affecting finite frequency properties \cite{HwangPlasmonPhonon,Ahn2014}, which are beyond the scope of the current work.

In general, developing a theory for electron-electron and electron-phonon interactions together, particularly when both interactions are strong, is a notoriously difficult and unsolved problem in condensed matter physics. Note that the interplay between electron-phonon and electron-electron interactions even in the Eliashberg theory of superconductivity remains essentially an unsolved hard problem for general values of the system parameters (e.g. when the Fermi velocity and phonon velocities are comparable as they are in tBLG for low twist angles).  The very successful so-called $\mu^*$ correction arising from electron-electron interaction in the Eliashberg theory is a drastic approximation, which applies very approximately only to metals where $T_F\gg T_D$, $E_F\gg \omega_D$,  and $v_F\gg v_{ph}$.  For tBLG, including both electron-electron and electron-phonon interactions on an equal footing in a general and quantitative strong-coupling theory is obviously hopelessly difficult task way beyond the scope of the current work.

For the acoustic phonon scattering induced electronic resistivity discussed in Sec.~\ref{sec:eph}, one possible correction arising from electron-electron interaction could, in principle, be the screening of the electron-phonon coupling by the free carriers themselves.  This issue has been discussed in depth with respect to regular monolayer and bilayer graphene (without any moir\'e flatband physics) in Ref.~\onlinecite{Min2011}.  Typically for metals, or more generally for situations involving electron energies much larger than the phonon energies involved in the scattering process (i.e. $E_F \gg \omega_D$), static screening approximation may be used to discuss the screening of the electron-phonon coupling by electron-electron interactions.  This leads to two main effects:  (1) The long wavelength acoustic phonon scattering, which is the main scattering process at low ($T \ll T_{BG}$) temperatures, is suppressed by screening leading to an effective decrease of the electron-phonon coupling compared to its bare value, i.e., the low-temperature BG regime has an effectively lower value of the deformation potential coupling because of static screening by the electrons themselves; and (2) because of the wavenumber dependence of static screening, an extra power of $q^2$, where $q$ is the wavenumber of the scattered phonon, comes into play inside the Bloch-Gr\"uneisen integral in Eqs.~(\ref{rhoTntheta}) and (\ref{rhoI}) of Sec.~\ref{sec:eph}, leading to the low-temperature BG resistivity temperature dependence changing to $T^{n+2}$ where $n$ is the BG resistivity power law ($n=4$ in 2D and 5 in 3D) without screening.  We emphasize that both of these screening induced modifications of phonon scattering happen only in the low temperature regime where long wavelength acoustic phonon are involved in the scattering process.  The higher temperature equipartition phonon scattering regime with the linear-in-$T$ resistivity behavior, which is the topic of our interest in the paper, is hardly affected by screening because the scattering in this linear-in-$T$ regime involves almost entirely large wavenumber $2k_F$ acoustic phonons which are essentially unscreened since screening occurs mostly at long distances (i.e. small wavenumbers).  Thus, even when screening is potentially important (i.e. when $E_F \gg \omega_D$), the interesting linear-in-$T$ resistivity produced by acoustic phonon scattering is hardly affected by screening, a fact which is not widely appreciated.  In addition, we point out that the screened BG behavior with a $T^6$ low-T dependence of the resistivity has never been observed in any 2D materials, so it is unclear that even the low-$T$ BG regime is actually affected by screening at all.  In fact, the experiment in Ref.~\onlinecite{Kim2010} clearly finds a $T^4$ (and not a $T^6$) power law dependence in the resistivity for $T \ll T_{BG}$ in very high density ($> 10^{13}$ cm$^{-2}$) graphene layers where, in principle, screening could play a role.  In addition, the experimental resistivity measured in Ref.~\onlinecite{Kim2010} is quantitatively consistent with a single graphene deformation potential electron-phonon coupling constant ($\sim $20 eV) throughout both low-$T$ BG ($\rho \sim T^4$) and high-T equipartition ($\rho \sim T$) regimes \cite{Hwang2008}. This experimental finding strongly argues against screening playing any role in graphene acoustic phonon scattering induced carrier resistivity even in the physical situation ($E_F \gg \Omega_D$ and $T \ll T_{BG}$) where screening might play a role.

The main physics of the interplay of electron-electron and electron-phonon interaction in the context of tBLG transport properties can only be captured by carrying out a fully dynamical theory where both electron-electron and electron-phonon interactions are frequency dependent.  Such a transport theory is highly demanding numerically since the relaxation time approximation no longer applies in such a dynamical situation, and all scattering must be treated as intrinsically inelastic, necessitating a full solution of the quantum transport integral equation (in both frequency and momentum) for the resistivity.  We believe that such a completely numerical theory is unwarranted for tBLG low-density ($< 2\times 10^{12} $ cm$^{-2}$)  transport since it is likely in the end to produce results very similar to the unscreened approximation in the linear-in-$T$ regime where phonon scattering effects are important.  The low-$T$ BG transport regime may indeed be affected by dynamical screening in some complicated nonuniversal and density-dependent manner, but this is of little practical significance since the phonon scattering contribution to the resistivity is strongly suppressed in this regime and the resistivity is likely to be dominated by other scattering processes for $T<5$ K any way.  We should comment, however, that a full dynamical treatment of screened electron-phonon interaction in the transport problem may even lead to an enhancement of the effective phonon scattering strength because it is well-known that dynamical screening could lead to anti-screening behavior \cite{DasSarma1988} because the dielectric screening function $\epsilon (q, \omega)$ behaves as $\epsilon (\omega)= 1 - \omega_p^2/\omega^2$ at high frequencies (where $\omega_p$ is the plasma frequency) which is always less than unity leading to a generic enhancement of the screened interaction.  By contrast, the static screening function for graphene $\epsilon (q, \omega=0) =1 + q_{TF}/q$, where $q_{TF}$ is the graphene Thomas-Fermi screening wavenumber\cite{DasSarma2011}, is always larger than unity leading to a suppression of the screened interaction.  Thus, whether screening leads to an enhanced or suppressed effective electron-phonon scattering strength is a subtle question requiring detailed calculations, which are beyond the scope of the current work.  We expect that our results and discussion in Sec.~\ref{sec:eph} with respect to phonon scattering limited $T$-linear resistivity being strongly enhanced by electron-phonon scattering in tBLG remains unaffected by electron-electron interactions.

In contrast to the above conclusion of electron-electron interaction not affecting our theory of phonon scattering effects in the tBLG carrier resistivity, we expect the electron-phonon interaction to affect the tBLG effective mass renormalization.  In fact, it is well-known that electron-phonon interaction enhances the thermodynamic effective mass as appearing, for example, in the specific heat.  This phonon-induced effective mass renormalization follows simply from the real part of the electronic self-energy due to electron-phonon coupling whereas the imaginary part of the same self-energy contributes to the resistivity (assuming no vertex corrections).  The phonon-induced effective mass renormalization (without considering any electron-electron interaction effects) itself is given in the leading order many-body perturbation theory by:
\begin{equation}
m^*/m= 1 + \lambda,
\end{equation}
where $m^*$ ($m$) are the renormalized (bare) effective masses, and  $\lambda$ is the dimensionless electron-phonon coupling strength which also enters the high-temperature resistivity through the formula:
\begin{equation}
\hbar /\tau= 2\pi \lambda  k_B T ,
\end{equation}
where $\tau$ is the scattering relaxation time entering into the Drude formula for the resistivity (with $N_0$ being the density of states at the Fermi energy):
\begin{equation}
1/\rho=e^2 v_F^{*2} N_0 \tau/2.
\end{equation}
The dimensionless tBLG electron-phonon coupling is proportional to the deformation potential coupling strength and is given by:
\begin{equation}
\lambda= N_0 D^2/(2 \rho_m v_{ph}^2).
\end{equation}
We note that the carrier density of states $N_0 \propto 1/v_F$, and hence $\lambda \propto 1/v_F \propto \alpha$, as mentioned already, leading to the strong enhancement of electron-phonon coupling in tBLG compared with ordinary MLG.
In small twist-angle tBLG, $\lambda \sim 1$ \cite{wu2019phonon} for $n \sim 10^{12}$ cm$^{-2}$ carrier density, leading to the phonon scattering rate manifesting the so-called Planckian behavior \cite{MIT2018_rho} with $\hbar/\tau \gg k_B T$ whereas the ordinary MLG and BLG have $\lambda(\sim 0.01) \ll 1$ leading to weak phonon-induced $T$-dependence in the resistivity even at room temperatures.

With respect to the effective mass renormalization, a $\lambda \sim 1$ corresponds to a factor of 2 increase in the renormalized effective mass over the bare mass arising in small angle tBLG purely from the strongly enhanced electron-phonon coupling whereas the corresponding electron-phonon renormalization in the MLG effective mass is of the order of $\sim$1\%.\cite{Tse2007}  Thus, the effective mass renormalization can indeed be strongly affected by electron-phonon interaction in addition to the electron-electron interaction effects discussed in Sec.~\ref{sec:ee}.  First, we note that the renormalization of the effective mass by electron-phonon interaction enhances the effective mass (i.e. it is a positive renormalization with $\lambda>0$) whereas the effective mass renormalization by the logarithmic flow of the electron-electron interaction effect is always negative with the renormalized effective mass being suppressed compared with the bare effective mass since the Fermi velocity is always enhanced as density decreases due to the RG flow.  In addition, the density of states for tBLG in the Dirac cone regime (i.e. low density) has a density dependence $N_0 \sim \sqrt{n}$, indicating that phonon-induced effective mass renormalization correction increases (decreases) with increasing (decreasing) carrier density.  Thus, the electron-phonon and electron-electron (Sec.~\ref{sec:ee}) interaction-induced effective mass renormalization behave the opposite ways (although both are proportional to the respective coupling constants $\lambda$ and $\alpha$).  The two effects can be distinguished, in principle, by careful measurements at low and high carrier densities (but staying within the Dirac cone approximation which implies $n<10^{12}$ cm$^{-2}$ always) where electron-electron and electron-phonon interactions would dominate respectively.  Whether such a separation of the two renormalization effects is experimentally viable through density-dependent effective mass measurements carried out in the regime (below the van Hove singularities) where the Dirac cone approximation is valid is unclear since the overall density regime is only about two decades in carrier density (at best $10^{10}$ to $10^{12}$ cm$^{-2}$).  

We note that the opposite effective mass renormalization correction of the electron-phonon interaction  compared with that of electron-electron interaction may be one of the underlying reasons for the experimental tBLG measurements finding very little mass or velocity renormalization as discussed in Sec.~\ref{sec:ee} --- the possibility of the two renormalization effects canceling each other out (at least within the large error bars of the currently available data) cannot be ruled out at this stage.  With more data over an extended density regime, one should be able to address this issue quantitatively since the density dependences of the two renormalizations are qualitatively different (i.e. logarithmic for the electron-electron interaction and square root for the electron-phonon interaction).

Another interesting question in this context is whether the two renormalization effects are additive as is often assumed in leading-order theories.  It may seem that the total effective mass renormalization arising from both electron-electron and electron-phonon interactions together, i.e. the experimental effective mass, can simply be written as (up to leading orders in the two interactions):
\begin{equation}
m^*/m= 1 + \lambda_{ep} + \lambda_{ee},
\label{eq:mstar}
\end{equation}
where $\lambda_{ep}$ equals $\lambda$ defined above and $\lambda_{ee}$ is the electron-electroon interaction dependent logarithmic effective mass renormalization discussed in Sec.~\ref{sec:ee}.  Note that both $\lambda_{ep}$ and $\lambda_{ee}$ include implied carrier density dependence not shown explicitly (i.e. subsumed in the $\lambda$).  Note also (as emphasized above) that the two renormalizations come with opposite signs-- $\lambda_{ep}$ ($\lambda_{ee}$) enhances (suppresses) the effective mass.  Although this additive renormalization appears reasonable in a theory involving leading order calculations for both interactions, it is not obvious at all that in the strong coupling situation, where both electron-phonon and electron-electron tBLG interactions are strong (by virtue of the strong flatband-induced suppression of the bare Fermi velocity for small twist angles), that the net renormalization would be additive.  It is entirely possible that because of the strong effects of the two interactions (and because they come with opposite signs), the theory would necessitate a Borel-Pad\'e resummation of the type discussed in Sec.~\ref{sec:ee} for the electron-electron interaction itself.  The issue of additive or not combined renormalization is beyond the scope of the current work and remains an interesting question for future investigations. The additive renormalization formula in Eq.~(\ref{eq:mstar}) is certainly correct within the leading-order theory in the two coupling constants, and hence may not be valid in the strong-coupling regimes of low twist-angle tBLG.  On the other hand, the fact that the two renormalizations come with opposite signs (i.e. electron-phonon and electron-electron interactions in general increase and decrease the effective mass of graphene, respectively) is quite general, and should apply in the strong-coupling regime also.  Therefore, it is possible that the strong electron-phonon interaction in tBLG serves to reduce the effect of electron-electron interaction-induced effective mass suppression, again bringing theory and experiment closer together.

\section{Conclusion}
\label{sec:con}
In summary, we have considered two aspects of many-body renormalization of low-density ($n \le 10^{12}$ cm$^{-2}$) tBLG electronic properties for low twist angle ($\theta<1.5^{\circ}$), where the continuum Dirac cone approximation should apply since the Fermi level is below the van Hove singularities. For the electron-phonon interaction, the small tBLG band velocity implies greatly enhanced ($\sim 1/v_F^{*2}$) electron-phonon coupling leading to very large and linear-in-$T$ resistivity for $T \ge T_{BG}/8$ where $T_{BG}=2\hbar v_{ph} k_F \propto \sqrt{n}$. The theory explains the available experimental data well for $T> 5$ K or so with the main discrepancy arising from the fact that a few samples at some specific densities appear to manifest linear-in-$T$ resistivity to temperatures almost an order of magnitude lower than that predicted by our theory (although most samples at most carrier densities agree well with the theory). We have proposed the possibility of the van Hove singularity driven Fermi surface Lifshitz transition and/or the gap opening at the Dirac point as possible reasons for the discrepancy, but much more experimental and theoretical work would be necessary to settle the question.

For the electron-electron interactions, we have investigated the role of the so-called ``coupling constant running'' in determining the low-density properties of tBLG, where the Dirac cone approximation applies, consequently leading to a logarithmic renormalization of the Fermi velocity and the relevant tBLG fine structure constant characterizing electron-electron interactions. The motivation here is that, given the large bare tBLG coupling constant $(\sim 1/v_F^*)$ arising from the small tBLG bare Fermi velocity, electron-electron interaction effects should be extremely strong in the tBLG low density ($<10^{12}$ cm$^{-2}$) regime according to the 1-loop perturbative RG theory. Carrying out a detailed comparison with the rather limited available tBLG experimental data on the measured Fermi velocity and effective mass, we conclude that the experimental observation of relative density independence of the measured low-density Fermi velocity is inconsistent with the 1-loop RG theory which predicts a large ($\sim$ by factors of $2-10$ depending on the twist angle) increase in the Fermi velocity as the carrier density decreases from $10^{12}$ cm$^{-2}$ to $10^{11}$ cm$^{-2}$. We propose that the tBLG electron-electron interaction effects are better described by a nonperturbative resummation theory and/or an $1/N$ expansion theory, both of which predict only a modest $(\sim 25\%)$ increase in the tBLG Fermi velocity, consistent with experimental findings, with decreasing density in the   $10^{11} - 10^{12}$ cm$^{-2}$ regime. 
An important finding of the current work is that the 1-loop RG theory fails to describe the many body effects in tBLG even near the charge neutrality point where the Dirac cone approximation applies because the low band velocity of the moir\'e flatband system makes the interaction strength large ($\gg 1$), where the applicable theory should be a strong-coupling theory (e.g. Borel-Pad\'e resummation, $1/N$ expansion) rather than a perturbative RG theory.  The fact that such strong-coupling theories do provide reasonable agreement between theory and experiment demonstrates that the continuum field theory is applicable to tBLG at low carrier energies below the van Hove singularity points.  One immediate implication of this strong-coupling behavior is that the effective mass renormalization is $\sim 25\%$ rather than being $\sim 500\%$ as the weak-coupling perturbative RG theory predicts.  A second implication is that Landau pole is absent in tBLG since the strong-coupling theories do not lead to a diverging running coupling at any finite energy.
We have also discussed qualitatively the interplay between electron-electron and electron-phonon interactions in tBLG, concluding that the carrier screening of the deformation potential coupling should play no role in the linear-in-$T$ equipartition regime of the tBLG electrical resistivity as well as that the effective mass renormalizations due to electron-electron and electron-phonon interactions may oppose each other at low carrier densities  possibly adding to the experimental null results on the velocity or mass renormalization effects although much more data would be needed before this cancellation can be validated.

One theory that remains to be developed for the future is a complete strong-coupling  Eliashberg theory for the electron-phonon interaction induced superconductivity in tBLG.  The weak-coupling theory predicts \cite{wu2019phonon} a superconducting $T_c \sim 1-10 $K in approximate agreement with experimental findings, but whether this result survives the strong-coupling limit or not can only be answered in the future when the full Eliashberg theory is developed for tBLG superconductivity.  One very serious challenge in the development of such an Eliashberg theory in tBLG is the fact that the Migdal approximation\cite{migdal1958interaction} is unlikely to apply to tBLG by virtue of the fact that $v_F\sim v_{ph}$ at low twist angles since the band velocity is strongly suppressed (see our Figs. \ref{Fig:vFDOS} and \ref{Fig:vFalpha}).  In the absence of the Migdal approximation, even the weak-coupling theory for superconductivity becomes a difficult problem, let alone the problem of strong-coupling superconductivity.  Note that the Migdal approximation is valid \cite{Bitan2014Migdal} in graphene as long as $v_F\gg v_{ph}$,  precisely the same condition necessary for the validity of Migdal theorem as in ordinary metals, however, tBLG violates this basic condition for the validity of Migdal approximation.

The main conclusion of this work is that the continuum field theories provide an excellent description of the low-density (and low-energy) properties of tBLG, when the electron-electron and electron-phonon interactions are both very strongly enhanced by virtue of the flatband Fermi velocity suppression in the moir\'e system, provided one uses the appropriate strong-coupling (either Borel-Pad\'e resummed RG theory or a strong-coupling  $1/N$ expansion) field theory for calculating the electron-electron interaction effects.  The 1-loop perturbative RG fails for tBLG since such a weak-coupling theory is inapplicable for $\alpha>1$. A corollary of our strong coupling theories is that graphene does not have any Landau pole.  The standard electron-phonon weak-coupling theory continues working well for the tBLG temperature-dependent resistivity since the dimensionless electron-phonon Eliashberg coupling remains small ($\lambda <1$) even near the magic twist angle in spite of its giant enhancement compared with untwisted graphene (where $\lambda \sim 0.001$).  We have indicated how these low-density theories can be extended to higher densities by explicitly incorporating the Lifshitz transition associated with the van Hove singularities in the density of states at higher energies where the Dirac cone approximation no longer applies.

\section{Acknowledgment}
This work is supported by the Laboratory for Physical Sciences.

\bibstyle{apsrev4-1} 
\bibliography{refs}

\end{document}